\documentclass[iop, revtex4]{emulateapj}
\usepackage{epstopdf}
\usepackage{multirow}
\usepackage{threeparttable}
\usepackage{natbib}
\usepackage{float}
\newcommand{\unit}[1]{\ensuremath{\, \mathrm{#1}}}

\newcommand{\Msol}{\hbox{$\mathrm{M_\odot}$}}
\newcommand{\msol}{\hbox{$\mathrm{M_\odot}$}}
\newcommand{\zfourge}{ZFOURGE}

\newcommand{\fq}{\hbox{$f_{q,sat}$}} 
\newcommand{\fqbg}{\hbox{$f_{q,bg}$}} 
\newcommand{\eq}{\hbox{$\epsilon_{q,sat}$}}

\newcommand{\myemail}{kawinwanichakij@physics.tamu.edu}
\slugcomment{Draft Version Nov 9, 2015}
\begin{document}
\title{Satellite Quenching and Galactic Conformity at $0.3<z<2.5$ 
\footnote{T\lowercase{his paper includes data gathered with the 6.5 meter} M\lowercase{agellan} T\lowercase{elescopes located at} L\lowercase{as} C\lowercase{ampanas} O\lowercase{bservatory}, C\lowercase{hile}.}}
\author{Lalitwadee Kawinwanichakij\altaffilmark{1$\dagger$}}
\author{Ryan F. Quadri\altaffilmark{1,2}}
\author{Casey Papovich\altaffilmark{1}} 
\author{Glenn G. Kacprzak\altaffilmark{3}}
\author{Ivo Labb\'{e}\altaffilmark{4}}
\author{Lee R.	Spitler\altaffilmark{5,6}}
\author{Caroline M. S. Straatman\altaffilmark{4}}
\author{Kim-Vy H. Tran\altaffilmark{1}}
\author{Rebecca Allen\altaffilmark{3,6}}
\author{Peter Behroozi\altaffilmark{7}}
\author{Michael Cowley\altaffilmark{5}}
\author{Avishai Dekel\altaffilmark{8}}
\author{Karl Glazebrook\altaffilmark{3}}
\author{W. G Hartley\altaffilmark{9,10}}
\author{Daniel D. Kelson\altaffilmark{11}}
\author{David C. Koo\altaffilmark{12}}
\author{Seong-Kook Lee\altaffilmark{13}}
\author{Yu Lu\altaffilmark{11}}
\author{Themiya Nanayakkara\altaffilmark{3}}
\author{S. Eric	Persson\altaffilmark{11}}
\author{Joel Primack\altaffilmark{12}}
\author{Vithal	Tilvi\altaffilmark{1}}
\author{Adam R. Tomczak\altaffilmark{1}}
\author{Pieter 	van Dokkum\altaffilmark{14}}


%
\altaffiltext{1}	{George P. and Cynthia W. Mitchell Institute for Fundamental Physics and Astronomy, Department of Physics and Astronomy,Texas A\&M University, College Station, TX 77843}						
\altaffiltext{2}	{Mitchell Astronomy Fellow}						
\altaffiltext{3}{Centre for Astrophysics and Supercomputing, Swinburne University, Hawthorn, VIC 3122, Australia}						

\altaffiltext{4}{Leiden Observatory, Leiden University, P.O. Box 9513, 2300 RA Leiden, The Netherlands}
\altaffiltext{5}{Department of Physics and Astronomy, Faculty of Sciences, Macquarie University, Sydney, NSW 2109, Australia}						
\altaffiltext{6}{Australian Astronomical Observatories, PO Box 915, North Ryde NSW 1670, Australia}						
\altaffiltext{7}{Space Telescope Science Institute, 3700 San Martin Drive, Baltimore, MD 21218, USA}							
\altaffiltext{8}{Center for Astrophysics and Planetary Science, Racah Institute of Physics, The Hebrew University, Jerusalem 91904, Israel}
\altaffiltext{9}{School of Physics and Astronomy, University of Nottingham, Nottingham NG7 2RD, UK}						
\altaffiltext{10}{Institute for Astronomy, ETH Zurich, Wolfgang-Pauli-Strasse 27, CH-8093 Zurich,Switzerland}
\altaffiltext{11}{The Observatories, The Carnegie Institution for Science, 813 Santa Barbara Street, Pasadena, CA 91101, USA}
\altaffiltext{12}{University of California Observatories/Lick Observatory, University of California, Santa Cruz, CA 95064, USA}
\altaffiltext{13}{Center for the Exploration of the Origin of the Universe, Department of Physics and Astronomy, Seoul National University, Seoul, Korea}

\altaffiltext{14}{Department of Astronomy, Yale University, New Haven, CT 06520, USA}		
\altaffiltext{$\dagger$}{\myemail}



\begin{abstract} 
 We measure the evolution of the quiescent fraction and quenching efficiency of satellites
around star-forming and quiescent central galaxies with stellar mass $\log(M_{\unit{cen}}/\msol)>10.5$ at $0.3<z<2.5$. We combine imaging from three deep near-infrared-selected surveys (ZFOURGE/CANDELS, UDS, and UltraVISTA),  which allows us to select a stellar-mass complete sample of satellites with $\log(M_{\unit{sat}}/\msol)>9.3$. Satellites for both star-forming and quiescent central galaxies have higher quiescent fractions compared to field galaxies matched in stellar mass at all redshifts.
We also observe ``galactic conformity'':  satellites around quiescent centrals are more likely to be quenched compared to the satellites around star-forming centrals.  In our sample, this conformity signal is significant at $\gtrsim3\sigma$ for $0.6<z<1.6$, whereas it is only weakly significant at $0.3<z<0.6$ and $1.6<z<2.5$. Therefore, conformity (and therefore satellite quenching) has been present for a significant fraction of the age of the universe.
%
%
The satellite quenching efficiency increases with increasing stellar mass of the central, but does not appear to depend on the stellar mass of the satellite to the mass limit of our sample.  When we compare the satellite quenching efficiency of star-forming centrals with  stellar masses 0.2 dex higher than quiescent centrals (which should account for any difference in halo mass), the conformity signal decreases, but remains statistically significant at $0.6<z<0.9$.
This is evidence that satellite quenching is connected to the star-formation properties of the central as well as to the mass of the halo.  We discuss physical effects that may contribute to galactic conformity, and emphasize that they must allow for continued star-formation in the central galaxy even as the satellites are quenched.
%
\end{abstract}

\section{Introduction}
\label{sec:intro}

Galaxies can be broadly classified as either quiescent or star-forming. As deep multiwavelength galaxy surveys have allowed us to obtain complete samples to higher and higher redshifts, it has become clear that a substantial population of quiescent galaxies exists out to at least $z \sim 4$  \citep[e.g.,][]{Cimatti2002,Bell2004,Papovich2006,Williams2009,Whitaker2011,Straatman2014}. However, the processes that are responsible for the quenching of star formation remain one of the central mysteries in the field of galaxy evolution. 

\par It has long been known that environmental processes act to inhibit star formation \citep[e.g.,][]{Dressler1980,Balogh1999,Kauffmann2004,Peng2010,Quadri2012,Kovac2014,Tal2014a}.  Although the exact mechanisms are not well-understood, it is generally expected that galaxies in dense environments (or more specifically, satellite galaxies) should lose their gas supply \citep{Gunn1972,Larson1980}. But quiescent galaxies are also found in low-density environments \citep{Kauffmann2004} and are often the central galaxy in their halo, in which case they will not be affected by satellite-specific processes. Thus there must be other ways to quench galaxies, and there has not been a shortage of proposed mechanisms: these include the shock-heating of infalling gas \citep{White1978,Dekel2006}, gas heating caused by minor mergers \citep{Johansson2009}, low-level AGN feedback \citep{Croton2006}, explosive AGN feedback \citep{Hopkins2006}, and the stabilization of gas disks \citep{Martig2009}. Finding clear observational evidence that either supports or rules out any one specific process has been notoriously difficult.

\par A new clue regarding galaxy quenching was presented by \cite{Weinmann2006}, who found that the star-formation activities of satellite and central galaxies at $z < 0.2$ are correlated. The correlation is such that the quiescent fraction of satellites is higher around quiescent central galaxies than around star-forming centrals. This phenomenon, which they refer to as ``galactic conformity", suggests that whatever process or processes cause the quenching of central galaxies also operate on their satellites.

\par Since the original \cite{Weinmann2006} result, a number of other studies have analyzed the correlation between the properties of satellites (i.e., specific star-formation rate, colors, and gas fraction) and their more massive centrals in the local universe using data from the Sloan Digital Sky Survey (SDSS). While \cite{Weinmann2006} refer to ``galactic conformity" as a correlation in the properties of central and satellite galaxies at fixed \emph{halo} mass, other studies have investigated conformity at fixed \emph{stellar} mass. These studies have generally confirmed that the quiescent fraction of satellites around quiescent centrals must be higher than those of star-forming centrals \citep[see][]{Ross2009,Kauffmann2010,Wang2012,Kauffmann2013,Knobel2014,Phillips2014,Phillips2015}.

\par However, there are a number of important questions that are raised by studies of galactic conformity at low redshift. One is, whether the satellites of star-forming centrals are quenched in excess of field galaxies at the same mass, or whether it is only the satellites of quiescent centrals that experience excess quenching. A second question is whether conformity exists only at fixed stellar mass, or whether residual signal is seen when the halo masses of the star-forming and quiescent centrals have been matched.
 
\citet{Wang2012} showed that the color distribution of satellites
is different for star-forming centrals than for quiescent centrals, consistent with galactic conformity. They also found that the satellites of intermediate-mass star-forming centrals are not
quenched at higher rates than mass-matched field galaxies, but that the satellites of higher mass star-forming centrals (with stellar masses $\log(M_{\unit{stellar}}/\Msol) > 11.1$) do show excess quenching.
 
\citet{Phillips2014} studied bright ($\sim0.1 L^{*}$) satellites around isolated $\sim L^{*}$ galaxies in SDSS at low redshift, $z < 0.032$. In order to narrow the range of halo masses probed by their sample, these authors required that central galaxies have exactly one bright satellite. They found that satellites of quiescent centrals are more likely to be quenched than stellar mass-matched field galaxies, but that satellites of star-forming centrals are similar to field galaxies (echoing the observational results of \citealt{Wang2012}). These authors also use the pairwise velocities between the centrals and satellites to show that the quiescent centrals occupy more massive halos than star-forming centrals. 

In a follow-up study, \cite{Phillips2015} also considered central galaxies with exactly two bright satellites. In this case the quenched fraction of satellites is nearly the same for star-forming and quiescent centrals, thereby reducing or eliminating the conformity signal. They also use the pairwise velocities to show that, for the systems with two satellites, the the halo masses of star-forming and quiescent centrals are consistent with each other. 

Taken together, the \citet{Phillips2014,Phillips2015} results suggest that conformity in the local Universe could be driven largely, or entirely, by a difference in halo mass between star forming and quiescent centrals.  This contrasts with conclusions based on SDSS group catalogs \citep[e.g.,][]{Weinmann2006,Knobel2014}. It is possible that studies based on group catalogs are affected by inaccurate halo mass estimates and by the misidentification of centrals and satellites, which can introduce a weak conformity signal at fixed halo mass even when none is present \citep{Campbell2015,Paranjape2015,Bray2015}. This leaves open the possibility that differences in the halo masses of quiescent and star-forming centrals are responsible for galactic conformity.
%
%

\par To study the physical cause of conformity, \citet{Wang2012} inspected mock catalogs from the \cite{Guo2011} semi-analytic model. They showed that within the model, the conformity effect can be partially
explained by the fact that quiescent centrals occupy more massive halos than star-forming centrals. But even at fixed halo mass, the
satellites of quiescent centrals were accreted at earlier (corresponding to the earlier overall assembly times of the parent halos) and were exposed to
more hot halo gas, which also contributes to conformity.

\par The findings of \citet{Kauffmann2013} point to an interesting addition to the idea of conformity as applying to central galaxies and their satellites. Using SDSS, they studied the correlation in star-formation activity between galaxies as a function of separation, including galaxies separated by small (intra-halo) and large (inter-halo) scales.  They found that the correlation depends on the stellar mass of the central:  for high-mass centrals, there is a correlation on small scales, within the dark-matter halo, consistent with previous observations of galactic conformity. For lower-mass centrals, \citeauthor{Kauffmann2013} found that a correlation in the star-formation properties of galaxies extends over many Mpc, beyond the putative virial radii of the individual galaxies. \citet{Kauffmann2015} also found that low-mass galaxies with low star formation rates have an excess of massive radio-loud neighbors extending to several Mpc. These results may be an indication that there are different processes at play, with a conformity effect present amongst the galaxies within a single parent dark matter halo and a separate effect acting on galaxies in neighboring halos \citep[but see][]{Paranjape2015}; these intra-halo and inter-halo effects have been dubbed ``1-halo" and ``2-halo" conformity, respectively.

Some insight into the physical cause of conformity may come from studying the evolution in the conformity signal with redshift, as this evolution depends on the underlying physics.  For example, there is some expectation that 1-halo conformity at low redshifts may be a result of 2-halo conformity at higher redshifts, because galaxies that are currently satellites were previously centrals in nearby halos. The 2-halo conformity in galaxy properties could be expected because of correlations in the recent or past assembly history of those halos, i.e. ``assembly bias" \citep[e.g.,][]{Gao2005,Croton2007,Tinker2008}. Recently \cite{Hearin2015} used the Bolshoi $N$-body simulation \citep{Klypin2011} to analyze the correlation between the mass-accretion rates of nearby halos. As the accretion rates of halos are correlated out to many times the halo virial radius, they argue that this effect may provide a physical basis for 2-halo conformity. \cite{Hearin2015} also predict that 2-halo conformity should be much weaker at higher redshifts ($z > 1$). If 1-halo conformity is entirely due to 2-halo conformity, this would suggest that 1-halo conformity should also disappear at $z > 1$.



In addition, there should be an evolutionary trend with redshift if conformity effects are associated with inter-halo effects.    Tinker \& Wetzel (2010) used clustering measurements with a halo-occupation distribution analysis to conclude that the evolution of the quenched fraction of satellites requires a quenching timescale that evolves with redshift as $T_Q \sim (1+z)^{-1.5}$, in the same way as the dynamical time, implying that the physical mechanism for satellite quenching should depend on the time that galaxies spend as satellites.    \cite{Wetzel2013} use $N$-body simulations combined with SDSS data to study satellite quenching as a function of both satellite and halo mass, and show that that SFRs for satellites are mostly unaffected for several Gyr after infall, but then they experience rapid quenching.  They further find that quenching timescales are shorter for more massive satellites, but do not depend on host halo mass because many satellites quench in lower-mass halos prior to infall. Therefore, key physical insight can be gained by studying the redshift evolution of satellite quenching. 


\par In this work we study the redshift evolution of galactic conformity on scales comparable to halo virial radii out to $z \sim 2.5$ (i.e.,~``1-halo" conformity between centrals and satellites). Our study is primarily concerned with galactic conformity at fixed stellar mass, but we also investigate the effects of halo mass. Previous studies have looked at the evolution of the correlation in star-formation activity of galaxies with environment (including within galaxy clusters) out to $z\sim 2$ \citep[see, e.g.,][and references therein]{Quadri2012,Bassett2013,Lee2015}, but have not generally studied conformity effects. To the best of our knowledge, the only comparable exploration of galactic conformity beyond the low-redshift universe was performed by \citet{Hartley2015}. They studied a sample of massive satellites (down to $M_\ast > 10^{9.7}~\msol$) around $\sim M^\ast$ central galaxies over $0.4 < z < 1.9$.   
They found evidence that galactic conformity persists over this redshift range, with higher quenched fractions of satellites 
around quiescent centrals compared to mass-matched samples of star-forming centrals.  Furthermore, they found 
that star-forming centrals have satellites with quenched fractions indistinguishable from field galaxies.   
\citet{Hartley2015} also argue that conformity is not simply due to a difference in halo mass between star-forming and quiescent centrals. This study was limited to a single field and to a smaller (and shallower) range in stellar mass.


\par Here, we use a new set of near-infrared (IR)-selected datasets, spanning multiple wide and deep fields to explore the correlation
between the star-formation activity of central galaxies and their satellites over a large range of stellar mass and $0.3 < z < 2.5$. The outline of this paper is as follows. In \S~2 we
describe our datasets and galaxy sample selection
criteria. In \S~3 we describe the method for identifying satellites
and for measuring the satellite quiescent fractions and quenching efficiencies.
 In \S~4 we explore how satellite quenching depends on the star-formation activity of the central galaxies, finding that conformity is present over our entire redshift range, although the statistical significance becomes weak beyond $z\sim1.6$. We also investigate satellite quenching as a function of the stellar mass of both centrals and satellites. In \S~5 we discuss these results, including the possible physical causes of conformity, and whether conformity persists at fixed halo mass. In \S~6 we
present our summary. Throughout, we define the process of ``galactic conformity" to be the correlation in star-formation activity between centrals and their satellites on scales comparable to the virial radius of the centrals' halos. With this definition our galactic conformity is akin to 1-halo conformity rather than 2-halo conformity. We adopt the following cosmological
parameters where appropriate, $H_0 = 70 \unit{km\;s^{-1}\;Mpc^{-1}}$,
$\Omega_{m} =0.3$, and $\Omega_{\Lambda} = 0.7$.  


\section{Data and Sample Selection}
\label{sec:data}
We select galaxies at $0.3 < z < 2.5$ from three datasets: the FourStar Galaxy 
Evolution Survey (ZFOURGE; PI Labb\'{e}), the UKIRT Infrared Deep 
Sky Survey (UKIDSS,~\citealt{Lawrence2007}) Ultra Deep Survey (UDS, Almaini et. 
al., in prep.), and the Ultra Deep Survey with the VISTA Telescope (UltraVISTA; \citealt{McCracken2012}).


\par We include galaxies at $0.3 < z < 1.6$ from a public 
$K_{s}$-selected catalog (Muzzin et al. 2013b) based on the first data release of 
UltraVISTA. The catalog covers a total area of 1.62 $\unit{deg}^{2}$ in the COSMOS
field \citep{Capak2007}. We construct our galaxy sample from the UltraVISTA by selecting galaxies with 
$K_s < 23$ mag, where the catalog is highly-complete.

\par In addition to UltraVISTA, at $0.3 < z < 1.6$, we also use the dataset which 
is based on UKIDSS UDS data release 8 \citep{Williams2009,Quadri2012}, the deepest degree-scale near-IR survey. The catalog covers an area of 0.65 
$\unit{deg}^{2}$, and the $K$-band reaches 24.6 mag (5$\sigma$ AB). To ensure a high level of completeness, we select a galaxy sample from this dataset with $K < 24$ mag.

At higher redshift, we draw our galaxy sample at $0.6 < z < 2.5$ from ZFOURGE (Straatman et al.\ 2015).
This survey is composed of three $11' \times 11'$
pointings with coverage in the CDFS \citep{Giacconi2002}, COSMOS, 
and UDS. The imaging reaches depths of $\sim 26 \unit{mag}$
in $J_{1}, J_{2}, J_{3}$ and $\sim 25 \unit{mag}$ in $H_{s}, H_{l}, K_{s}$
\citep[see][Straatman et al. 2015]{Spitler2012,Tilvi2013,Papovich2015}.
The medium-band filters from ZFOURGE provide an advantage by sampling the Balmer break at $1<z<4$ 
better than broadband filters alone. As in \citet{Kawinwanichakij2014}, we combine the 
ZFOURGE data with public HST/WFC3 F160W and F125W imaging from 
CANDELS \citep{Grogin2011,Koekemoer2011} in the three
fields. As described by \cite{Tomczak2014}, we make use of the CANDELS F160W image as the 
detection band to select a sample of galaxies at $z<2.5$ to low masses ($10^{9.3}~\msol$). 

\par We rederive photometric redshifts, rest-frame colors and stellar masses for the public 
UDS and UltraVista catalogs using the same method as for our ZFOURGE catalogs to
ensure as homogeneous a dataset as possible.   
Photometric redshifts and rest-frame colors are derived using EAZY 
\citep{Brammer2008}. We use the default set of spectral
templates derived from the PEGASE models \citep{Fioc1997} and a dust
reddened template derived from the \cite{Maraston2005} model to fit
the $0.3-8$~$\mu$m photometry for each galaxy to obtain its
photometric redshift, but note that the templates are iteratively tweaked during the fitting process.  Similarly, we derive stellar masses using
\cite{Bruzual2003} stellar population models with FAST code
\citep{Kriek2009}, assuming exponentially declining star formation
histories, solar metallicity, and a \cite{Chabrier2003} initial mass
function.

For our study, the relative redshift errors between the centrals and satellites are paramount,
and traditional photometric redshift testing (comparing photometric to spectroscopic redshifts) 
is infeasible as the satellite galaxies in our sample are typically much fainter than 
spectroscopic magnitude limits. We estimate the relative uncertainties in photometric redshifts 
between the centrals and satellites using the technique described by \citet{Quadri2010}, in which the photometric redshift differences in close galaxy pairs are measured. Since many close galaxy pairs are physically-associated, each galaxy provides an independent 
estimate of the true redshift.  Therefore, the distribution of the
differences in the photometric redshifts of galaxy pairs can be used
to estimate the photometric redshift uncertainties. 
For ZFOURGE, the typical photometric redshift
uncertainties at $1 < z < 2.5$ in the COSMOS, CDFS, and UDS
fields are $\sigma_z$= 0.06, 0.07, and 0.08, respectively (where $\sigma_z =\sigma /\sqrt{2}$, 
and where $\sigma$ is the width measured from a Gaussian fit to the distribution of pair redshift differences
in each field, and the $\sqrt{2}$ accounts for the fact that we take the difference between two independent measurements).  
For the UDS we derive $\sigma_z = 0.05$ and 0.04 for galaxies at $0.5 < z < 1.0$ and $1.0 < z < 1.5$, respectively.  
For UltraVISTA, we derive $\sigma_z=0.01$ and 0.05 for the same redshift ranges.

\par We explore the evolution of satellite quenching over $0.3 < z < 2.5$ by dividing our galaxy sample into four redshift bins,
each spanning roughly the same interval of cosmic time ($1.4-2.3$ Gyr): 
$0.3 < z < 0.6$, $0.6 < z < 0.9$, $0.9 < z < 1.6$, and $1.6 < z < 2.5$. 
To guard against possible survey-to-survey systematic biases, we select galaxy 
samples from at least two surveys depending on the stellar mass-completeness limit, as explained below. 
But for the highest redshift bin we can only use ZFOURGE because the UDS and UltraVISTA datasets are not deep enough to identify satellites to our desired mass range.

\par In this paper we consider central galaxies 
and their satellites, which are defined in \S~\ref{sec:samplesel} and 
\S~\ref{sec:selsat}. We denote the stellar masses of the centrals as 
$M_\mathrm{cen}$ and the stellar masses of the satellites as $M_{\mathrm{sat}}$. 
We use $f_{q,sat}$ and $\epsilon_{q,sat}$ to denote the quiescent 
fraction and quenching efficiency of satellite galaxies.
Those quantities have been corrected for projected background galaxies
using the same method as in \citet[and see below]{Kawinwanichakij2014}.

\vspace{20pt}
\begin{deluxetable}{cccc}
\tabletypesize{\footnotesize}
\tablecolumns{4} 
\tablewidth{0pt}
 \tablecaption{ Stellar mass completeness limits for three datasets at $0.3 < z < 2.5$
 \label{table:masscompletetable}}
 \tablehead{
 \colhead{} \vspace{-0.1cm}& \colhead{UDS} &  \colhead{UltraVISTA}  & \colhead{ZFOURGE}  \\ \\ 
  \colhead {Redshift} & {$\log(M(z))$} & {$\log(M(z))$} & {$\log(M(z))$}       \\ 
   \colhead{} \vspace{-0.1cm}& {$\log(M_\ast / \mathrm{M}_\odot)$} &  {$\log(M_\ast / \mathrm{M}_\odot)$}   & {$\log(M_\ast / \mathrm{M}_\odot)$}   \\ \\ 
  }
 \startdata 
 0.3& \phn8.3& \phn8.7& 7.7\\
 0.4&\phn 8.5& \phn8.9& 7.9\\
0.5& \phn 8.7& \phn9.1& 8.1\\
0.6&\phn 8.9& \phn9.3& 8.3\\
0.7&\phn 9.0& \phn9.4& 8.4\\
0.8& \phn9.2& \phn9.6& 8.6\\
0.9& \phn9.3& \phn9.7& 8.7\\
1.0&\phn 9.4& \phn9.8& 8.8\\
1.1&\phn 9.5& \phn9.9& 8.8\\
1.2&\phn 9.6& 10.0& 8.9\\
1.3& \phn9.7&10.1& 9.0\\
1.4&\phn 9.7&10.1& 9.0\\
1.5& \phn9.8&10.2& 9.1\\
1.6&\phn 9.9&10.3& 9.2\\
1.7& 10.0&10.4& 9.2\\
1.8&  10.0&10.4& 9.3\\
1.9& 10.1&10.5& 9.3\\
2.0& 10.2&10.6& 9.3\\
2.1& 10.2&10.6& 9.4\\
2.2& 10.3&10.7& 9.4\\
2.3& 10.3&10.7& 9.4\\
2.4& 10.4&10.8& 9.5\\
2.5& 10.4&10.8& 9.5\\
 \enddata
\vspace{-0.3cm}

\end{deluxetable}

\subsection {Stellar Mass-Completeness}
\label{sec:masscomplete}
\begin{figure*}
\epsscale{1.0}
\plotone{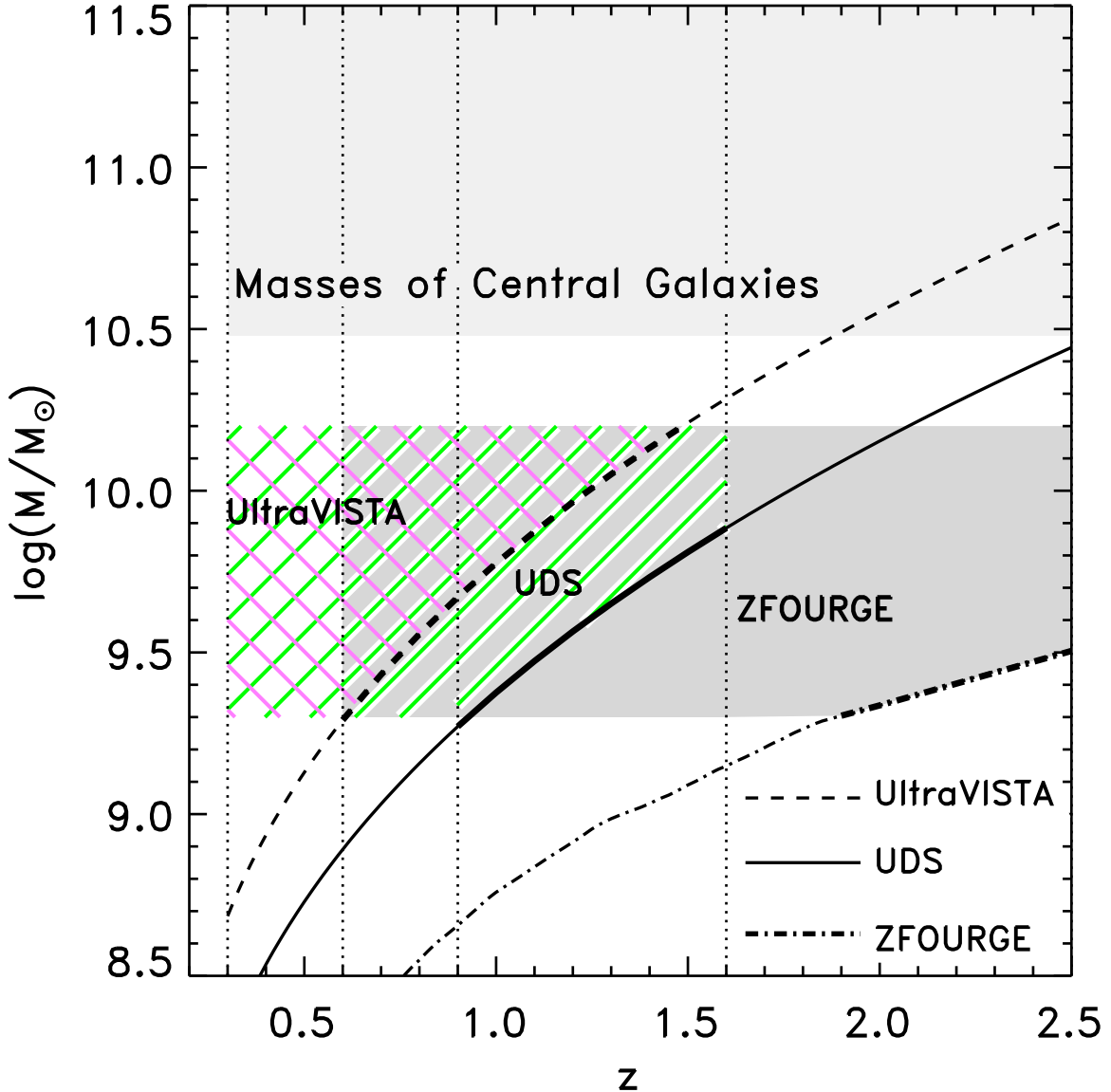}
\caption{95\% Stellar mass-completeness limit vs.\ redshift computed for quiescent 
galaxies in three datasets: UltraVISTA ($K_{s} < 23$ mag; dash curve), 
UKIDSS UDS ($K < 24$ mag; solid curve), and ZFOURGE ($H_{160} < 26.5$ mag; dot dash curve). 
The mass-completeness limits for UDS and UltraVISTA are derived using the technique described 
by Quadri et al. (2012), whereas the mass-completeness limits for
ZFOURGE is determined from passively evolving a SSP with a formation redshift $z_{f} = 5$. 
The light grey shaded region shows the stellar mass ranges of our samples centrals.
The dotted vertical lines indicate our redshift bins used in this study. The thicker curves 
show the redshift and stellar mass ranges where we count neighboring galaxies using 
lower mass limit that evolves with redshift. The green and magenta diagonal hatched
regions show the stellar mass ranges of satellites from UDS and UltraVISTA, whereas the gray shaded region is for ZFOURGE.}
\label{fig:masscomplete}
\end{figure*}

\par Understanding the stellar mass-completeness limit for each dataset is crucial for our analysis. Because we are concerned with the galaxy quiescent fractions, it is important that we are highly-complete for both star-forming and quiescent galaxies. Quiescent galaxies have higher mass-to-light ratios, and therefore we adopt 95\% mass-completeness limits for galaxies
with quiescent stellar populations. In Figure~\ref{fig:masscomplete}, 
we plot the adopted stellar mass-completeness 
limits for galaxies from ZFOURGE, UDS, and UltraVISTA at $0.3 < z < 2.5$. For UDS and UltraVISTA, we employ an updated version of the technique described by \cite{Quadri2012} to 
estimate the 95\% mass-completeness limit that corresponds to the magnitude limit as a function of 
redshift. We select quiescent galaxies in narrow redshift bins, scale their fluxes and masses 
downward until they have the same magnitude as our adopted limit $K=24.0$ for UDS 
and $K_{s} = 23.0$ for UltraVISTA. Then we define the mass-completeness limit as 
the stellar mass at which we detect 95\% of the dimmed galaxies at each redshift.

The empirical technique to derive stellar mass completeness (used for UltraVISTA and UDS) may be inaccuate for ZFOURGE. The ZFOURGE catalogs were selected using a different bandpass (WFC3/$H_{160}$ for ZFOURGE compared to $K$ for the other fields), so we are unable to scale directly the mass limits determined from those surveys to ZFOURGE. Additionally, the estimates of the 95\% mass completeness limits in ZFOURGE may be inaccurate using the empirical method because the smaller ZFOURGE fields do not allow for a precise determination of the mass-to-flux ratio distributions of quiescent galaxies in narrow redshift bins.

Therefore, for ZFOURGE we determined the stellar mass-completeness limits  using a stellar population synthesis model  \citep[using EzGal,][]{Mancone2012} for a passively evolving single stellar population with a \citet{Chabrier2003} IMF, solar metallicity, a formation redshift $z_{f} = 5$, and $H_{160} < 26.5$ mag. This gives a slightly higher (i.e., more conservative) stellar mass completeness limit than what we would have derived using the empirical method (which we used for UltraVISTA and UDS). Moreover, it could be argued that one should use a lower formation redshift for lower mass galaxies because observationally lower mass galaxies have lower mass-to-light ratios \citep[see, e.g.,][]{Speagle2014}.   However, we use the conservative assumption of $z_f=5$ in order to ensure that our sample of low-mass quiescent galaxies is highly complete even for galaxies with the highest stellar-mass--to--light ratios.

We provide the adopted completeness limits for UDS, UltraVISTA, and ZFOURGE at $0.3 < z < 2.5$ in Table~\ref{table:masscompletetable}.

\begin{deluxetable*}{ccccc}
\tabletypesize{\footnotesize}
\tablecolumns{5} 
\tablewidth{0pt}
 \tablecaption{ Number of quiescent centrals and star-forming centrals in three datasets at $0.3 < z < 2.5$
 \label{table:sampleno}}
 \tablehead{
 \colhead{Stellar mass range} \vspace{-0.1cm}& \colhead{Redshift} &  \colhead{Dataset} & \colhead{$N_{\unit{c}}(\mathrm{Quiescent})$} & \colhead{$N_{\unit{c}}(\mathrm{Star-forming})$}\\ 
\vspace{0.1cm}} 
 \startdata 
 \cutinhead{\bf{Central mass:} $\log(M_{\mathrm{cen}}/\Msol) > 10.5$}

           \bf{Satellite mass:}  $\log(M_{\mathrm{sat}}/\Msol) = 9.3-9.8$   &  $ 0.3 < z < 0.6 $ & UDS           &263 & 134 \\ 
                     &                             & UltraVISTA & 846  & 701\\ \\
              
           \cline{2-5}     
                      &                           &                    &         &      \\
                     &  $ 0.6 < z < 0.9 $& UDS           & 468  & 317 \\ 
                     &				 & UltraVISTA &  1494 & 1375 \\
                     &                             & ZFOURGE   & 92  & 91  \\ \\
             \cline{2-5}  
              &                           &                    &         &      \\
                      & $ 0.9 < z < 1.6 $& UDS 		& 1207& 1486 \\
                      &                           & UltraVISTA   & 2770 & 3924 \\
                      &                           & ZFOURGE   & 156   & 219  \\ \\
               \cline{2-5}  
                      &                           &                    &         &      \\
                      & $ 1.6 < z < 2.5 $& ZFOURGE      & 140  & 199  \\
                    &                           	&                   &         &  \\ 
      \bf{Satellite mass:}  $\log(M_{\mathrm{sat}}/\Msol) = 9.8-10.2$ &  $ 0.3 < z < 0.6 $&UDS    & 263 &  134\\ 
                     &                             & UltraVISTA & 846 & 701\\ \\
                \cline{2-5}
                 &                           &                    &         &      \\
                     &  $ 0.6 < z < 0.9 $& UDS          & 468 & 317\\ 
                  &                             & UltraVISTA   &  1494 & 1375 \\
                     &                             & ZFOURGE     &  92 & 91 \\ \\
                \cline{2-5}
                 &                           &                    &         &      \\
                      & $ 0.9 < z < 1.6 $& UDS 		& 1207 & 1486 \\
                      &                           & UltraVISTA   & 2770 & 3924\\
                      &                           & ZFOURGE   & 156   &  219 \\ \\
                 \cline{2-5}
                  &                           &                    &         &      \\
                     & $ 1.6 < z < 2.5 $& ZFOURGE      & 140  & 199  \\
   \cutinhead{\bf{Central mass:} $10.5 < \log(M_{\mathrm{cen}}/\Msol) < 10.8$}

      \bf{Satellite Mass:}  $\log(M_{\mathrm{sat}}/\Msol) = 9.3-10.2$ &  $ 0.3 < z < 0.6 $&UDS          & 161&  108 \\ 
          
                     &                             & UltraVISTA & 369 & 479\\ \\
                     \cline{2-5}
                 &                           &                    &         &      \\
                   &  $ 0.6 < z < 0.9 $& UDS           & 288 & 240 \\ 
                       &		         & UltraVISTA & 762  & 951  \\
                     &                             & ZFOURGE   & 38 & 53 \\ \\
                     \cline{2-5}
                 &                           &                    &         &      \\
                      & $ 0.9 < z < 1.6 $& UDS 		& 656 & 977 \\ 
                       &                           & UltraVISTA  & 1461 & 2652 \\
                      &  			& ZFOURGE   & 79  & 120  \\ \\
                      \cline{2-5}
                 &                           &                    &         &      \\
                      & $ 1.6 < z < 2.5 $& ZFOURGE      & 71 & 95   \\
       \cutinhead{\bf{Central mass:} $\log(M_{\mathrm{cen}}/\Msol) > 10.8$}

       \bf{Satellite mass:}  $\log(M_{\mathrm{sat}}/\Msol) = 9.3-10.2$  &  $ 0.3 < z < 0.6 $&UDS          & 102 &  26\\
                            &                             & UltraVISTA & 477 & 222\\ \\
                            \cline{2-5}
                 &                           &                    &         &      \\
                     &  $ 0.6 < z < 0.9 $& UDS           & 180 & 77 \\ 
                     &			         & UltraVISTA &  732 & 424 \\
                     &                             & ZFOURGE   &  53 & 38 \\ \\
                     \cline{2-5}
                 &                           &                    &         &      \\
                      & $ 0.9 < z < 1.6 $& UDS 		& 551 & 509 \\
                      & 			& UltraVISTA  	& 1309 & 1272 \\
                      & 		          &ZFOURGE   & 77  & 96  \\	 \\
                      \cline{2-5}
                 &                           &                    &         &      \\
         & $ 1.6 < z < 2.5 $& ZFOURGE      & 65  & 100  \\

 \enddata
\vspace{-0.3cm}

\end{deluxetable*}
\subsection{Selection of Centrals and $UVJ$ Classification}
\label{sec:samplesel}
\par Our goal is to measure the fraction of quiescent satellites ($f_{q,sat}$) around massive
galaxies at $0.3 < z < 2.5$. We select central galaxies from the three datasets
with $\log(M_{\mathrm{cen}}/\Msol) > 10.5$ (i.e., $M_\mathrm{cen}  > 3 \times 10^{10}$~\msol). We also study the dependence of 
satellite quenching on the stellar mass of central galaxies, and will consider subsamples of central galaxies with $10.5 <
\log (M_\mathrm{cen}/\msol) < 10.8$ and $\log (M_\mathrm{cen}/\msol) > 10.8$ (i.e., $> 6 \times 10^{10}~\msol$).
A summary of number of centrals from each 
galaxy sample is given in Table ~\ref{table:sampleno}. 

\par Similar works by \cite{Tal2013} and \cite{Hartley2015} applied isolation criteria for the selection of 
central galaxies. They considered galaxies as ``central'' if no other, more massive galaxies are 
found within a projected radius of 500 pkpc (proper kpc). Otherwise, galaxies are counted as satellites 
of their more massive neighbors. \citet{Phillips2014} applied a similar isolation criteria for 
galaxies with $\log(M_{\mathrm{cen}}/\Msol) > 10.5$ by allowing no other galaxies with 
similar stellar mass within a projected distance
of 350 pkpc. In addition to this isolation criterion, they allow no more than one 
galaxy with $\log(M_{\mathrm{cen}}/\Msol) > 10.5$ within an inner (outer) radius of 350 pkpc (1 pMpc).

We apply a similar rejection criterion for our central galaxy sample selection, as contamination in our sample of centrals can potentially introduce a spurious conformity signal \citep{Campbell2015}. We exclude galaxies 
from our sample of centrals if there is a more massive galaxy within a projected radius of 
300 ckpc (comoving kpc). 
We opt to use this comoving aperture size as it is approximately the virial 
radius of a halo with mass of $\log(M/\Msol) \sim 12.0$ over our redshift range, 
which is near the halo mass of our intermediate-mass galaxy sample ($10.5 < \log (M_\mathrm{cen}/\msol) < 10.8$). 
Because the virial radius increases weakly with halo mass, 
we also test an isolation criteria of 500 ckpc.  We find that the conformity signals described in \S~\ref{sec:satquenching} persist, but the significance decreases
because the sample size drops by $20-50\%$, 
so we adopt 300 ckpc isolation criteria.


\begin{figure*}
\epsscale{1.0}
\plottwo{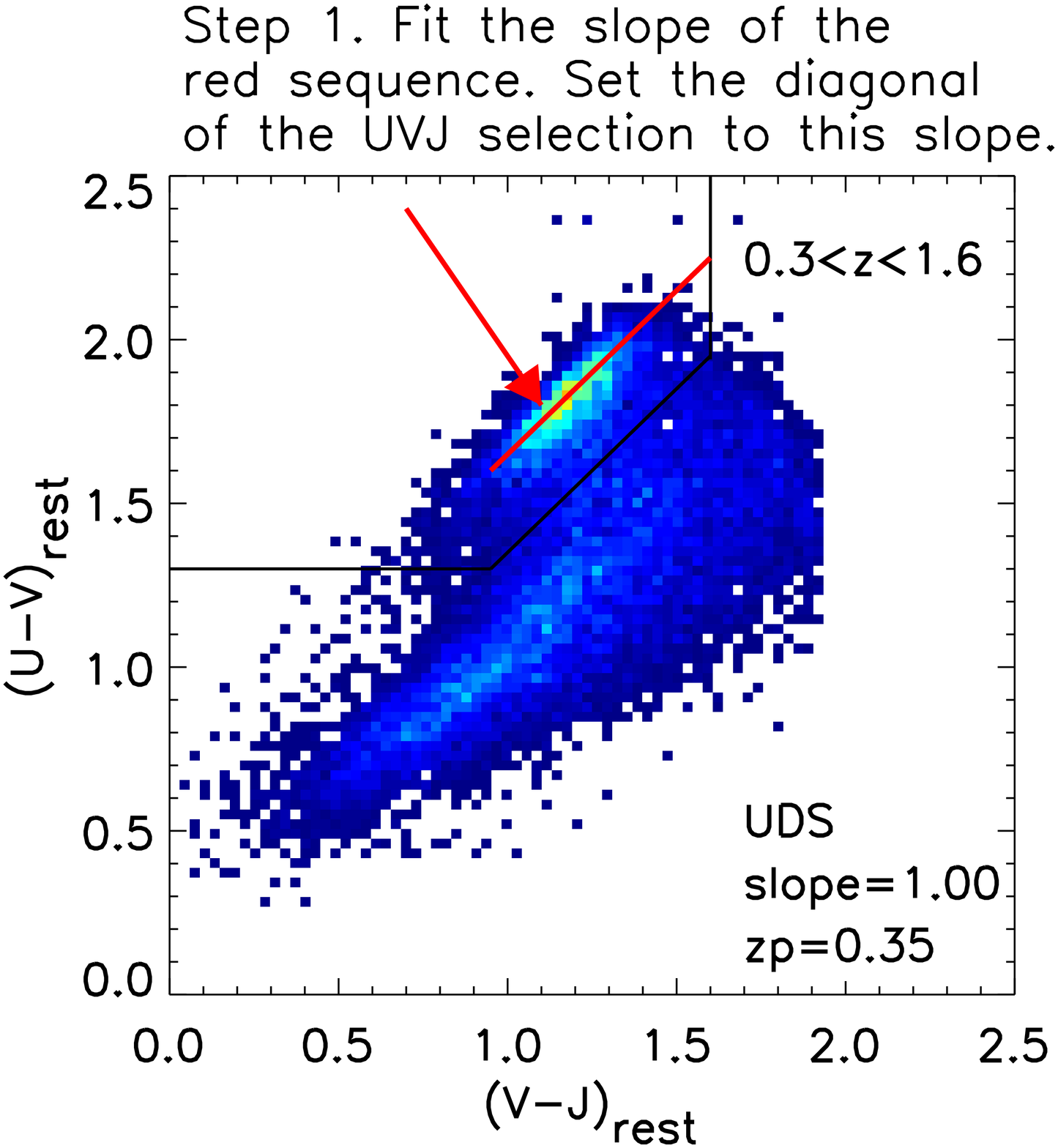}{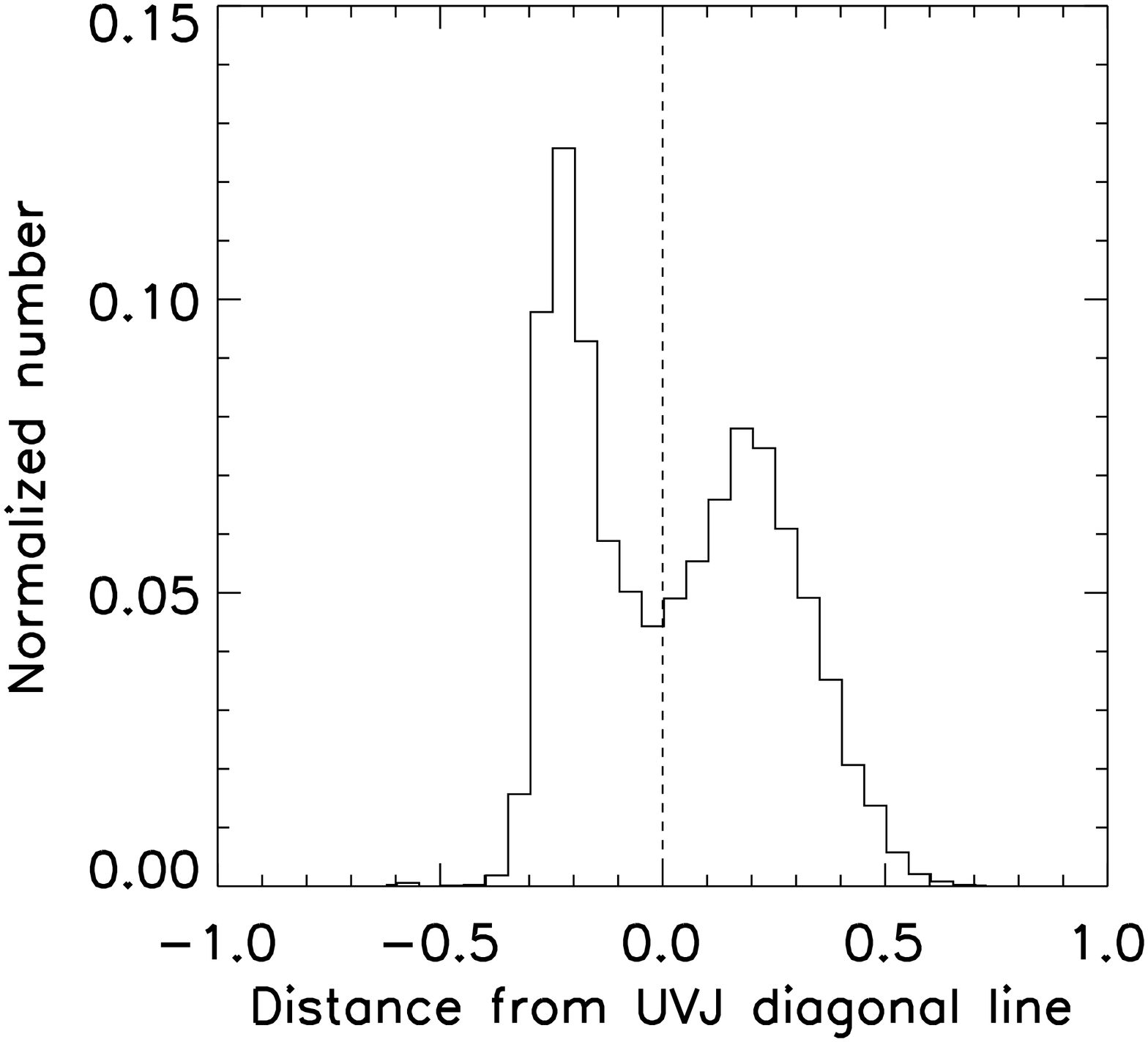}
\plottwo{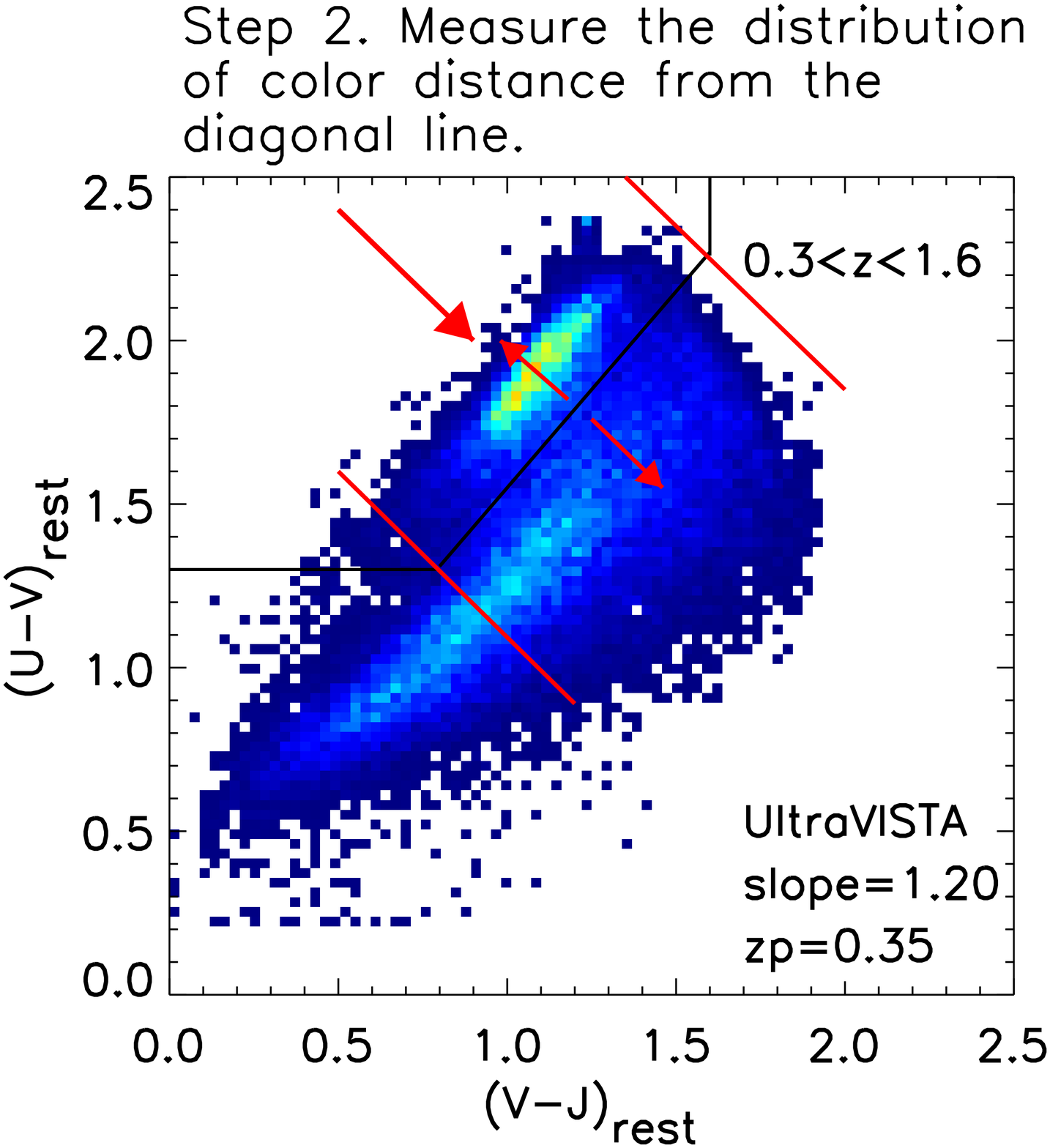}{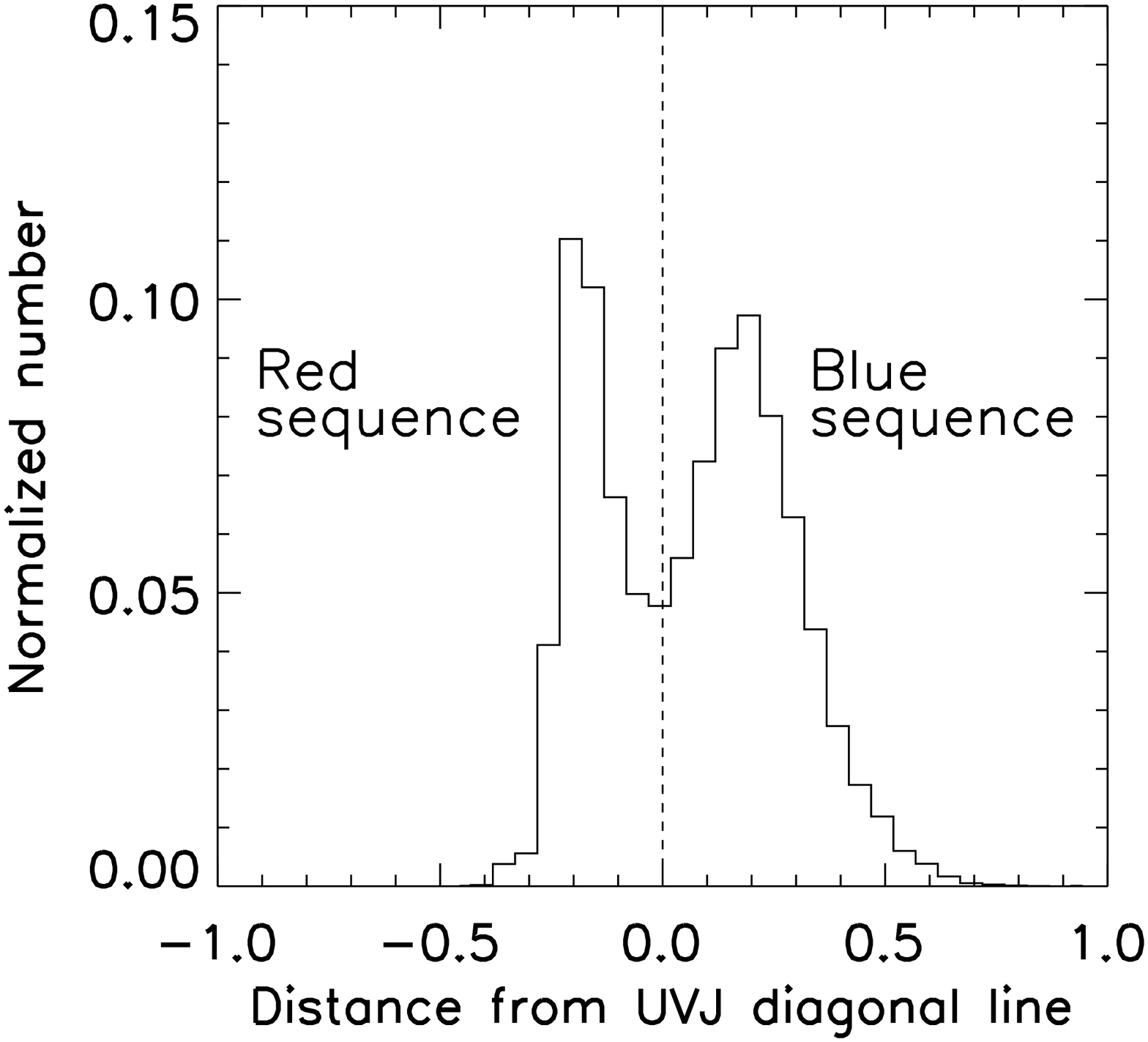}
\plottwo{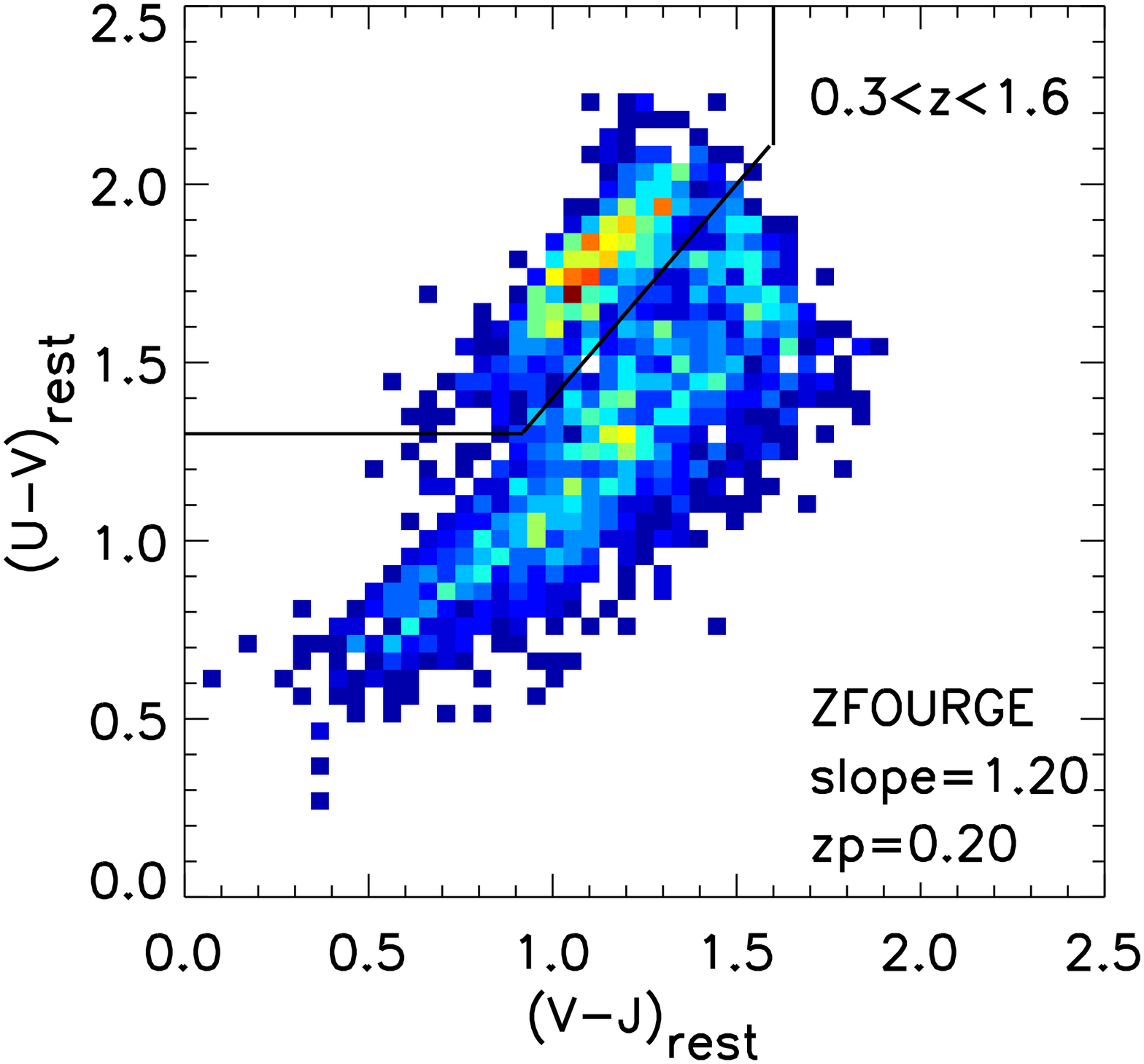}{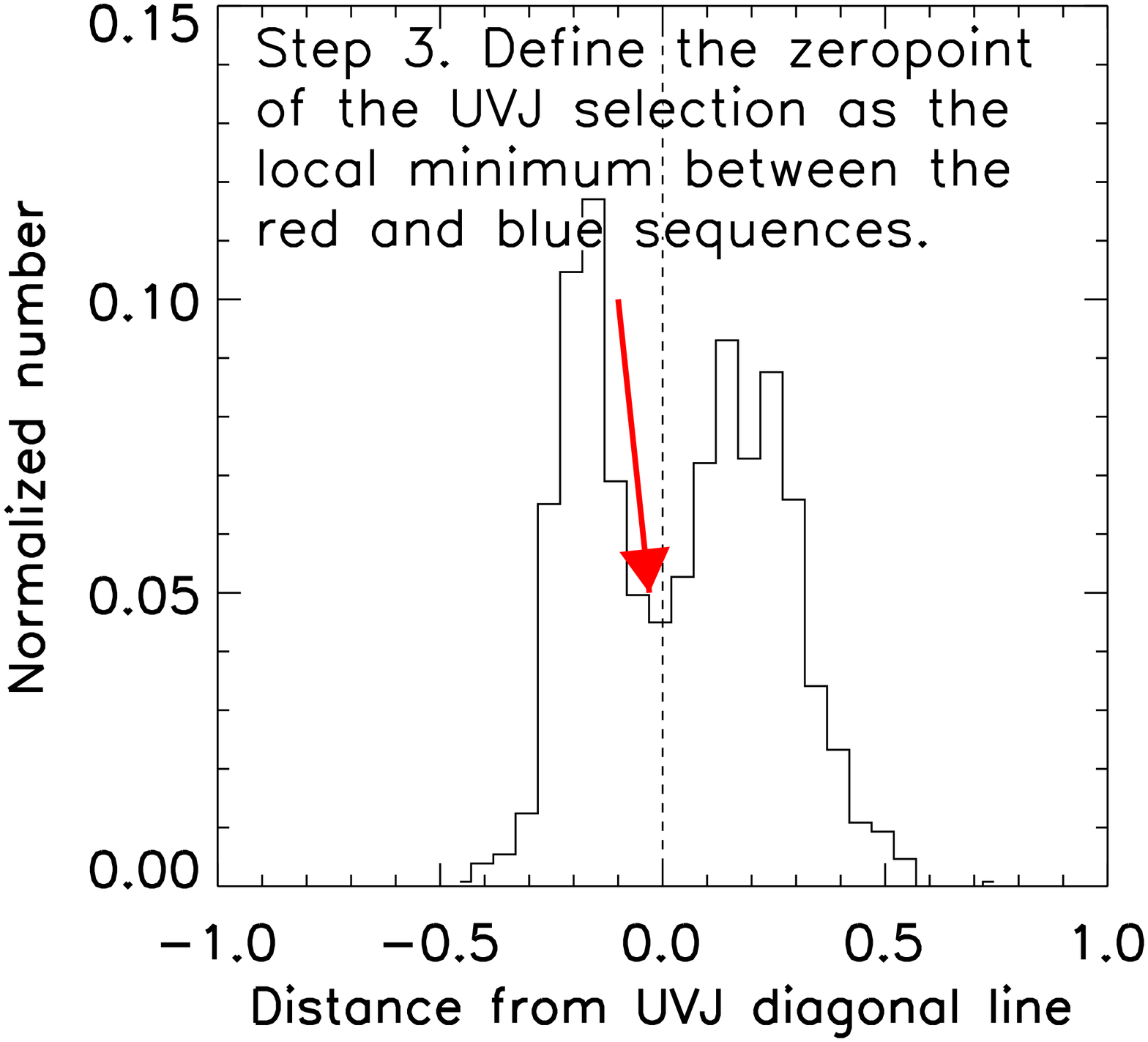}
\caption{\emph{Left}: Rest-frame $U-V$ versus $V-J$ color 
for galaxy sample with $\log(M_*/\Msol)>9.8$ at $0.3 < z < 1.6$. The galaxies in the upper left region of the plot (separated by the solid line) 
are quiescent; galaxies outside this region are star forming. \emph{Right}: 
Distribution of the distance (in color) from the diagonal line in $UVJ$ color  
(the slope $A$, see Equation 1) that separates the quiescent and star-forming 
sequences in the $UVJ$ color space.  
We define the zeropoint of the $UVJ$ quiescent region as the local minimum in this distribution, 
indicated by the vertical dashed line.}
\label{fig:uvjdiagram}
\end{figure*}

\par We classify galaxies as either star-forming or quiescent based on the rest-frame $U-V$ versus $V-J$ color-color diagram \citep[$UVJ$ diagram; e.g.,][]{Williams2009,Whitaker2011}. Our early tests of the different catalogs showed that there exist (small) systematic variations in the rest-frame colors of galaxies at fixed mass and redshift in different surveys.  To remove the effect of these systematic variations on our analysis,
we implement a method to self-calibrate the region delineating 
the colors of star-forming and quiescent in the color-color space (Figure~\ref{fig:uvjdiagram}).
We start by defining a generic region of the $UVJ$ diagram for quiescent galaxies as, 

 \begin{eqnarray} \label{eq:uvj}
  U-V &>& A \times (V-J) +zp     \nonumber \\
  U-V &>& 1.3 \nonumber \\
   V-J &<&1.6
\end{eqnarray}

\noindent where $A$ and $zp$ are variables we derive as follows. We fit for $A$ as the slope of the red
sequence in the $UVJ$ plane, finding slopes of $A=1.2, 1.0$, and 1.2 for ZFOURGE, UDS, 
and UltraVISTA, respectively. Next, we measure the distribution of the distance in $UVJ$ color 
from the diagonal line defined by the slope $A$  in Equation~1 
(where the ``color distance'' is the distance in $UVJ$ color from the line). We measure the zeropoint $zp$ as the local minimum
between the two peaks in the $UVJ$ color distribution. Figure~\ref{fig:uvjdiagram} 
shows a demonstration. We find zeropoints of $zp=0.2, 0.35$, and 0.35 
for ZFOURGE, UDS, and UltraVISTA. Our method therefore removes any systematics
between the data and/or in the analysis of the survey catalogs and minimizes any differences
in the definition of the quiescent region for the $UVJ$ diagram.
Table~\ref{table:sampleno} gives the numbers of quiescent and star-forming galaxies in the UDS, UltraVISTA,
 and ZFOURGE fields.

\subsection{Selection of Satellites}
\label{sec:selsat}
  
To identify satellites of the central galaxies in our sample, 
we first select all neighboring galaxies around each central
from our sample that satisfy the following satellite conditions
\begin{eqnarray}
\left| z_\mathrm{cen} - z_\mathrm{sat} \right| \leq 0.2 \nonumber \\
M_\mathrm{lim}\ \leq M_\mathrm{sat} < 10^{10.2}~\Msol
  \label{eq:criteria}
\end{eqnarray}

\noindent where $z_\mathrm{cen}$ and $z_\mathrm{sat}$ are the
photometric redshift of the central and satellite, respectively.
$M_\mathrm{sat}$ is the stellar mass of
the satellite, and $M_\mathrm{lim}$ is the lower-mass 
limit, which is shown in Figure~\ref{fig:masscomplete}.
Our requirement that
$\Delta z = |z_\mathrm{cen} - z_\mathrm{sat}| \leq 0.2$ is motivated by
our relative photometric uncertainty ($\sigma_z$) between centrals
and satellites as mentioned in \S~\ref{sec:data}.  
In each case, the $\sigma_z$ values for
galaxies are less than half the $\Delta
z \leq 0.2$ requirement in Equation~2, which argues that this
selection criterion is appropriate. The stellar mass limits for satellites
we study is $9.3 < \log(M_{\mathrm{sat}}/\Msol) < 10.2$, and later
we subdivide this into bins of $9.3 < \log(M_{\mathrm{sat}}/\Msol) < 9.8$ and
$9.8< \log(M_{\mathrm{sat}}/\Msol) < 10.2$ in order to test for variations in the quenching 
efficiency as a function of satellite mass. 
\par Our primary results in this paper are determined using an evolving
stellar-mass limit, in which we only select satellites in each field
that are above the mass-completeness limits (See \S~\ref{sec:masscomplete} and Figure~\ref{fig:masscomplete}).
This maximizes our sample size and boosts the significance of our results.
For example, at $0.6 < z < 0.9$ and $0.9 < z < 1.6$, where the UltraVISTA galaxy sample starts to become incomplete,
 we then only use satellites from that survey lying above the mass completeness curve
(shown as the hatched region above the thick dash curve in Figure~\ref{fig:masscomplete}). In principle this may affect our results, since in some of the redshift/mass bins the mean redshift and the mean satellite stellar mass will differ slightly between our fields. However this is a small effect as the satellite quenching efficiency does not depend strongly on satellite mass (\S~\ref{sec:dependenceonsatmass}), and moreover we have verified that none of our main results change if we use fixed lower mass limit at all redshifts ($\log(M_{\mathrm{sat}}/\Msol) > 9.3$).

\section{Environmental quenching of satellite galaxies}
\label{sec:method}

\subsection{Identifying Satellites using Statistical Background Subtraction}

\par To perform a statistical analysis of the average quiescent
fraction of satellites around our sample of massive galaxies, we use a
statistical background subtraction technique
\citep[e.g.,][]{Kauffmann2010,Tal2012,Wang2012,Kawinwanichakij2014}. We
detect objects within fixed apertures centered on our central galaxies and
satisfying Equation~\ref{eq:criteria}. These apertures include both
physically-associated galaxies as well as chance alignments of
foreground and background galaxies. We estimate and correct for the
contamination due to chance alignments by placing random apertures
across the field. We adapt this procedure by restricting the
placement of the random apertures to regions near to the centrals, as
demonstrated by \cite{Chen2006}.  This accounts for the bias due to
contaminating galaxies that are physically-associated with the
centrals, but are not satellites (i.e.,\ the two-halo term of the correlation
function; see \citealt{Chen2006})\footnote{The contaminating galaxies that are physically-associated with the central galaxies in our sample are expected to have marginally different properties than truly random field galaxies due to the fact that they exist in biased regions of the Universe. There may be an additional effect due to large-scale 2-halo conformity. If 2-halo conformity exists, our procedure effectively corrects for it.}. We therefore place the
random apertures within annuli with inner and outer radii equal to 1 and 3
cMpc from each central galaxy for the UDS and UltraVISTA. 
Parenthetically, our tests showed that the restriction on the location of the background apertures has 
only a small effect on the conformity signal. 
Relative to apertures that are placed randomly through the field, this correction
increases the quiescent fractions of background galaxies by 0.4\%--10\%.
For the smaller ZFOURGE fields, placing the random apertures within annuli is too restrictive, and for this survey we randomly place the apertures across the fields. We do note that the ZFOURGE fields are small enough that even these randomly-placed apertures trace the same large-scale environment as
the centrals. Additionally, we find that when we restrict the
background apertures to be $>3$ cMpc from the centrals, 
it changes the measured quenching efficiencies (see \S~\ref{sec:satquenching} below)
by ~10\%, and none of our conclusions would be changed.

Both the random and real apertures have a radius of 300 \unit{ckpc}.
We experimented using 300 pkpc apertures (i.e.,~apertures with a fixed physical
size rather than fixed comoving size), and find that our main
conclusions are not appreciably affected by the choice of aperture.
We therefore adopt the measurement of quiescent fraction within a
circular aperture of 300 ckpc for the rest of this paper. 
We also tested a plausible range of aperture sizes, and found they do not appreciably change the results. In the Appendix we show the effect on the quenching efficiencies of satellites around quiescent and star-forming centrals using these different-sized apertures (both comoving and physical aperture radii).

\subsection{Matching the Stellar Mass Distribution of Star-Forming and Quiescent Central Galaxies}
\label{sec:matchmass}
Quiescent galaxies have a stellar mass distribution that is shifted to higher stellar masses compared to star-forming galaxies.
 Therefore, any observation that satellites around quiescent central galaxies may be preferentially 
 quenched may be caused by a difference in the stellar mass of the centrals.  
 Therefore we match the stellar mass distributions of the quiescent 
 and star-forming central galaxies.  
Following the method of \cite{Hartley2015} we construct a histogram of stellar masses of 
central galaxies in bins of $\Delta \log(M_{\mathrm{cen}}) = 0.1$ and use this to calculate a weighting factor for each stellar mass bin of quiescent centrals ($w_{i}^{q}$) using
\begin{eqnarray}
w_{i}^{q} = \frac{N_{\mathrm{cen},i} }{N_{\mathrm{cen},i}^{q}} 
\end{eqnarray}
Similarly, we calculate the weighting factor for each stellar mass bin of star-forming centrals ($w_{i}^{sf}$) using
\begin{eqnarray}
w_{i}^{sf} = \frac{N_{\mathrm{cen},i} }{N_{\mathrm{cen},i}^{sf}} 
\end{eqnarray}
where $N_{cen,i}$ is the total number of central galaxies in stellar mass bin $i$ and
$N_{\mathrm{cen},i}^{q(sf)}$ is the number of quiescent (star-forming) centrals in stellar mass bin $i$.

\par In each bin of central stellar mass, we weight the number of satellites by  $w_{i}^{q}$ for quiescent centrals and by $w_{i}^{sf}$ for star-forming centrals. This effectively matches the stellar mass distributions of both the quiescent and star-forming
 centrals to the stellar mass distribution of all central galaxies
 \citep[this is similar to the method used to match the stellar mass distributions of centrals in][]{Kawinwanichakij2014}.

\par In addition to the difference in the stellar mass distributions, there are slight differences in the redshift distributions of quiescent and star-forming centrals within each redshift bin. For example, if,  at fixed stellar mass,  the star-forming galaxies tend to lie at higher redshift, then this could possibly affect our results. However, we argue this is not the case.  In each redshift bin, the difference in mean redshift between the star-forming and quiescent centrals is small, and comparable to the photometric redshift uncertainty, $\Delta z \lesssim 0.02 (1+z)$. Furthermore, if differences in the redshift distributions of the star-forming and quiescent centrals were important, we would expect the quenching efficiency of star-forming galaxies to be more similar to the quenching efficiency of quiescent galaxies in adjacent redshift bins. As we show below (\S~4.1), this is not the case: the quenching efficiency of satellites around quiescent galaxies is consistently higher than that for star-forming centrals in any of the other redshift bins at $0.6 < z < 0.9$, $0.9 < z < 1.6$, and $1.6 < z < 2.5$. Therefore, it seems unlikely that the (small) differences in the redshift distributions of the quiescent and star-forming centrals within each redshift bin contribute significantly to the observed galactic conformity signal.

\subsection{Average Quiescent Fraction}
\label{sec:qfrac}

\label{sec:qfrac4surveys}
\begin{figure*}
\epsscale{1.0}
\plottwo{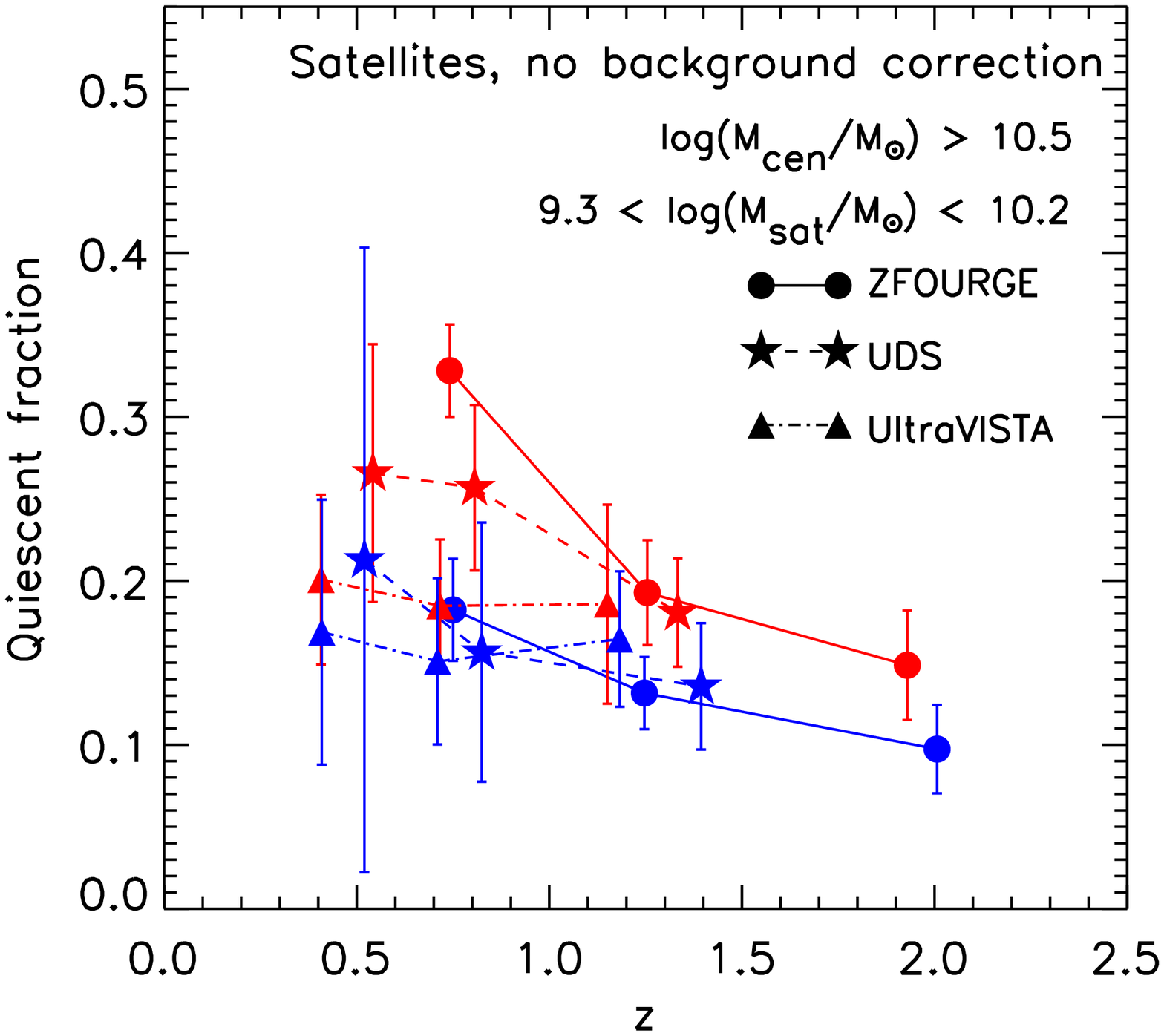}
{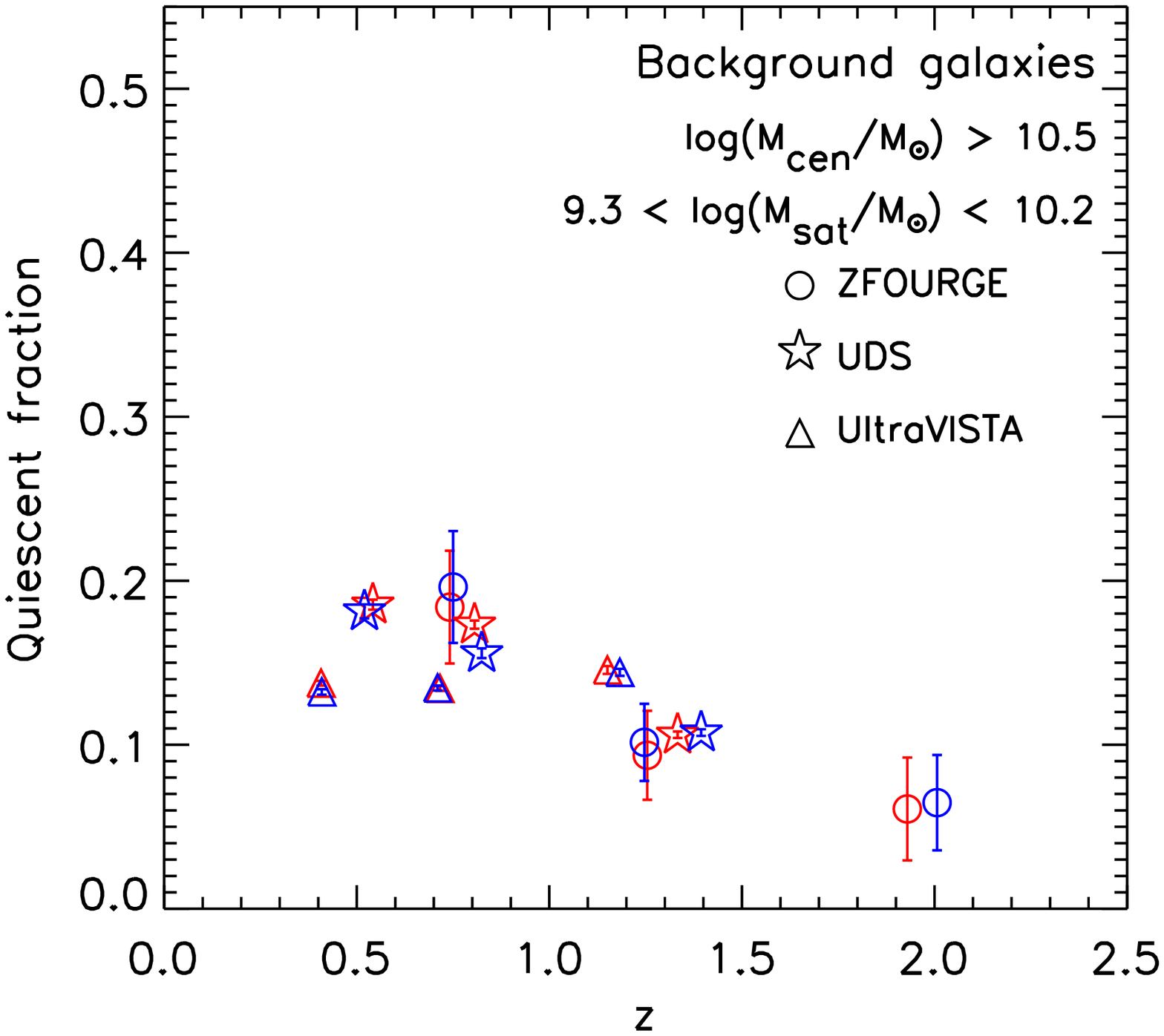} 
\caption{\emph{Left:}  The evolution of the average quiescent fraction of neighboring 
quiescent galaxies ($f_{q,nei}$) with stellar mass of $9.3 < \log (M_\mathrm{sat}/\msol) < 10.2$ 
around the quiescent centrals (red) and the star-forming centrals (blue) 
($\log(M_{\unit{cen}}/\Msol)  > 10.5$) from three datasets: ZFOURGE (circles), 
UKIDSS UDS (stars), and UltraVISTA (triangles). 
\emph{Right:} Same as the left panel but for the 
average quiescent fraction of neighboring background ($f_{q,bg}$) measured in random apertures. 
We use the measurement in random apertures to account for physically associated galaxies
as well as chance alignment of foreground and background galaxies (\S~\ref{sec:method}).
The error bars are based on the 68 percentile of the distribution of the  quiescent fraction of satellites from a
bootstrap resampling technique (\S~\ref{sec:errorestimate}). The UltraVISTA and UDS points have been offset to lower and higher redshift slightly for clarity.} 
\label{fig:qfracrealrandom}
\end{figure*}

\par We count the number of quiescent and star-forming neighboring galaxies
in apertures around central galaxies in redshift bins. We define ``neighboring galaxies''
as those in the vicinity of the centrals that satisfy the Equation~\ref{eq:criteria} 
(neighboring galaxies include both satellites and foreground or background objects along the line of sight).
The quiescent fractions of neighboring galaxies 
($f_{q,nei}$) are shown in the left panel of Figure~\ref{fig:qfracrealrandom}.
We then perform the same measurement with the random apertures. As shown in the right panel 
of Figure~\ref{fig:qfracrealrandom}, the quiescent fractions of galaxies 
measured in random apertures (\fqbg) tend to be lower than for the neighboring galaxies, and are quite consistent among the surveys, 
with $\left\langle \fqbg \right\rangle \sim 0.2$.

We estimate the average quiescent fraction of satellites ($f_{q,sat}$) using 

\begin{eqnarray}
f_{q,sat}=\frac{\Sigma(N_{nei}^{q}-N_{bg}^{q})}{\Sigma(N_{nei}^{tot}-N_{bg}^{tot})}
\label{eq:fqsat}
\end{eqnarray}
\noindent where $N_{nei}^{q}$ and $N_{nei}^{tot}$ are the number of neighboring quiescent galaxies 
and the total neighboring galaxies, respectively, around a central. Similarly,
$N_{bg}^{q}$ and $N_{bg}^{tot}$ are the number of neighboring quiescent galaxies and 
the total neighboring galaxies, respectively, measured in the random aperture. 
The summation is for all central galaxies in a given subsample of stellar mass and/or redshift.
The resulting fraction ($f_{q,sat}$) represents the average fraction of quiescent satellites 
around that sample of central galaxies.
\subsection{Average Quenching Efficiency}
\label{sec:qeff}

In this work we are concerned with the difference in quiescent fractions of satellites and
background galaxies. This difference, normalized by the star-forming fraction of the background
galaxies, gives a direct estimate of the fraction of satellites that have been quenched in excess 
of the quenched field galaxy population,

\begin{eqnarray}
\epsilon_{q,sat} =\frac{f_{q,sat}-f_{q,bg}}{1-f_{q,bg}}
\label{eq:eqsat}
\end{eqnarray}
where $f_{q,sat}$ is the quiescent fraction of satellites measured around centrals,
and $f_{q,bg}$ is the quiescent fraction of satellites measured in random apertures.
We refer to $\epsilon_{q,sat}$ as the quenching efficiency.


\subsection{Error estimation}
 \label{sec:errorestimate}
 We estimate the uncertainty on the quiescent fraction (\fq) and the quenching efficiency (\eq)
measurements using a bootstrap resampling technique. We generate 100,000 bootstrap samples for each 
subsample of quiescent and star-forming centrals. We then measure the satellite quiescent fractions and the quenching 
efficiencies for each set of bootstrap samples. We calculate the uncertainty as the 68 percentile of the distribution 
of the  quiescent fraction (or quenching efficiency) of satellites from the bootstrap samples.
The error bars estimated from these bootstrap resamplings are up to 3 times larger than the Poisson uncertainties.

\par We also use the uncertainties from a bootstrap technique of each field and survey (the three ZFOURGE fields, UDS, and UltraVISTA) to calculate weights for combining the results from the fields. We use this combined dataset for our analysis, but we also discuss survey-to-survey variations.

\subsection{Significance estimation}
 \label{sec:signiestimate}
It is desirable to assign a significance statistic ($p$--value) when
comparing the differences between the quiescent fraction of satellites
(or the quenching efficiency of satellites) for different subsamples.
We estimate the significance as the fraction of bootstrap samples
(\S~\ref{sec:errorestimate}) in which the
quiescent fraction (or the quenching
efficiency) of satellites around star-forming centrals is equal
or greater than that of quiescent centrals. We denote the
$p$--value derived from the bootstrap resampling technique as
$p$.

\section{Dependence of Satellite Quenching on Galaxy Properties}
\label{sec:satquenching}

\subsection{The Detection of Satellite Quenching and Galactic Conformity to $z\sim2$}

\begin{figure}
\epsscale{1.0}
\plotone{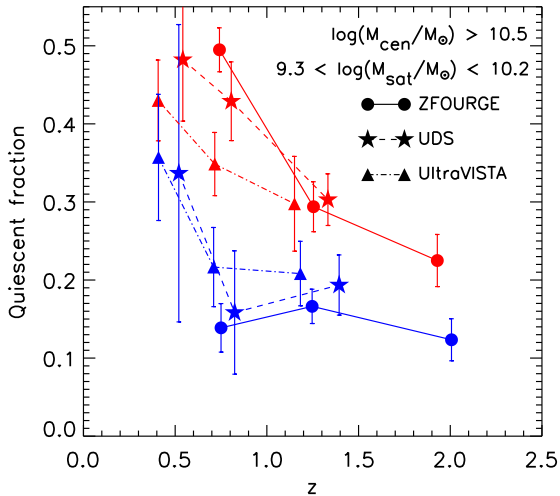}
\caption{The evolution of the average quiescent fraction ($f_{q,sat}$)
of satellites with stellar mass of $9.3 < \log (M_\mathrm{sat}/\msol) < 10.2$ 
around quiescent centrals (red) and star-forming centrals (blue) 
($\log(M_{\unit{cen}}/\Msol)  > 10.5$) from three datasets: ZFOURGE (circles), 
UKIDSS UDS (stars), and UltraVISTA (triangles). 
The error bars are based on the 68 percentile of the distribution 
of the  quiescent fraction of satellites from the bootstrap samples.
  For all fields and redshift ranges, we see evidence for higher quiescent fractions for satellites around quiescent centrals
compared to satellites around star-forming centrals at fixed stellar mass.
The UltraVISTA and UDS points have been offset to lower and higher redshift slightly for clarity.} 
\label{fig:qfrac4surveys}
\end{figure}

 \begin{figure}
\epsscale{1.0}
\plotone{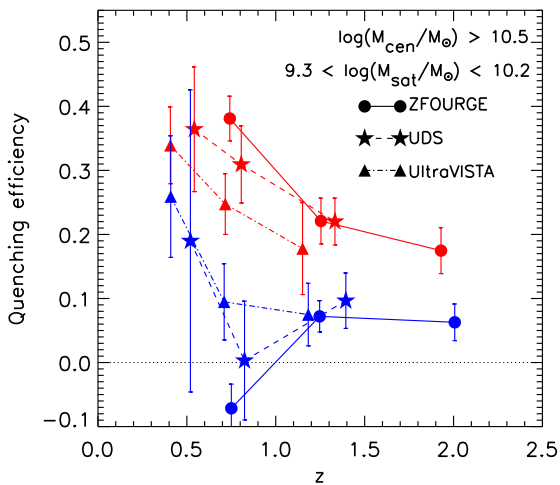}
\caption{The average quenching efficiency of satellites (\eq) with stellar mass of 
$9.3 < \log (M_\mathrm{sat}/\msol) < 10.2$ around central galaxies 
($\log(M_{\unit{cen}}/\Msol)  > 10.5$) from three datasets: ZFOURGE (circles), 
UKIDSS UDS (stars), and UltraVISTA (triangles).
The horizontal dotted line at  $\epsilon_{q,sat} = 0$ indicates where a galaxy 
is not quenched as it becomes a satellite of a central galaxy. 
The quenching efficiency of satellites around quiescent centrals is higher compared 
to those around the star-forming centrals, although the effect is most pronounced at 
$0.6 < z < 0.9$. The positive quenching efficiency of satellites of star-forming centrals 
(at least at $z < 0.6$) indicates that satellites of star-forming centrals are more quenched 
compared to background galaxies at the same stellar mass.
The UltraVISTA and UDS points have been offset to lower and higher redshift slightly for clarity.}
\label{fig:qeff4surveys}
\end{figure}

\begin{figure*}
\epsscale{1.0}
\plotone{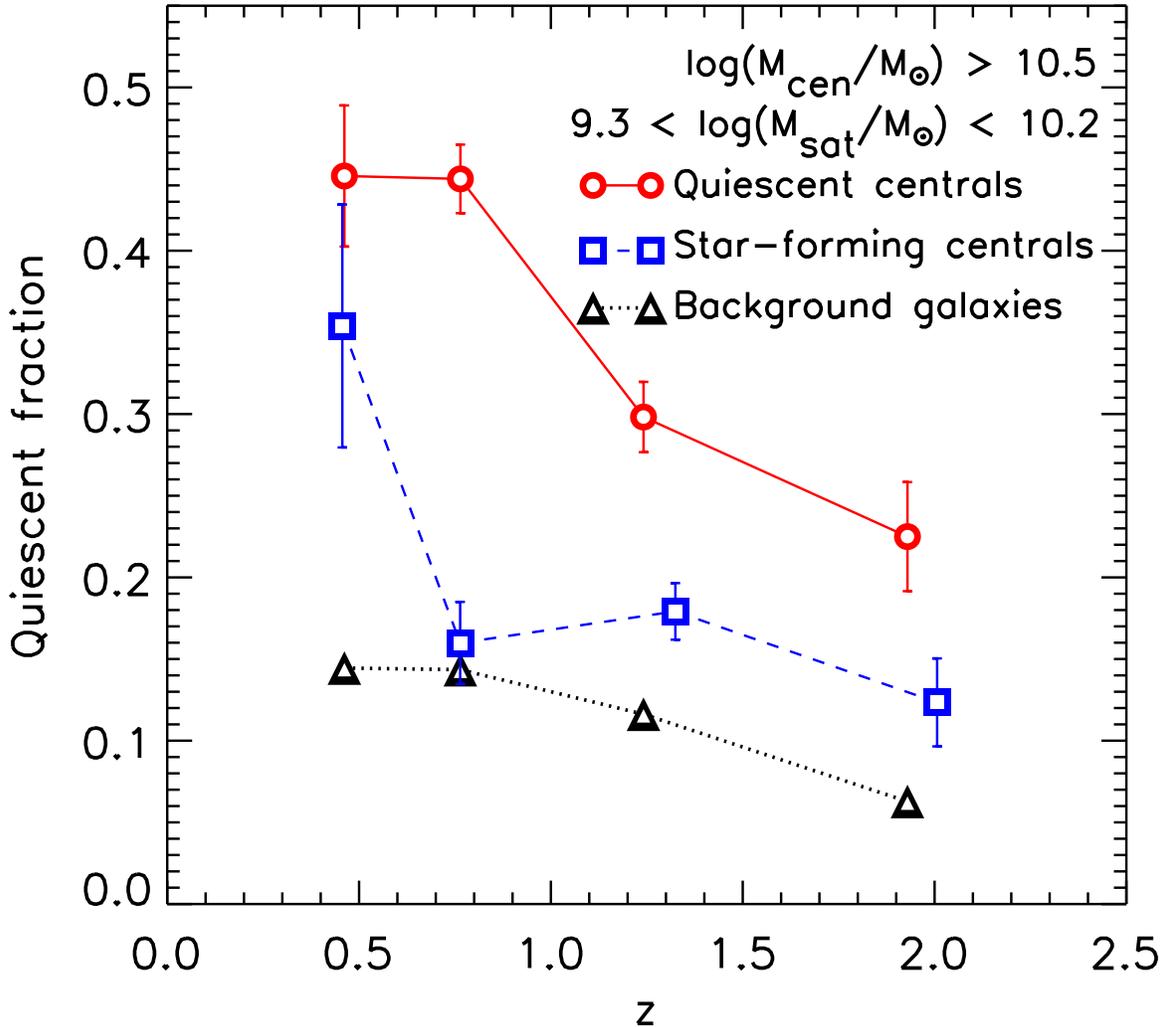}
\caption{The average quiescent fraction of satellites (\fq) with stellar mass of 
$9.3 < \log (M_\mathrm{sat}/\msol) < 10.2$ around quiescent centrals (red circles) and
star-forming centrals (blue squares) with stellar mass of $\log(M_{\unit{cen}}/\Msol)  > 10.5$
combining from the three datasets. The average quiescent fraction of background galaxies 
of the same stellar masses are also shown (black triangles).
}
\label{fig:qfraccomballmsat}
\end{figure*}

\begin{figure*}
\epsscale{1.0}
\plotone{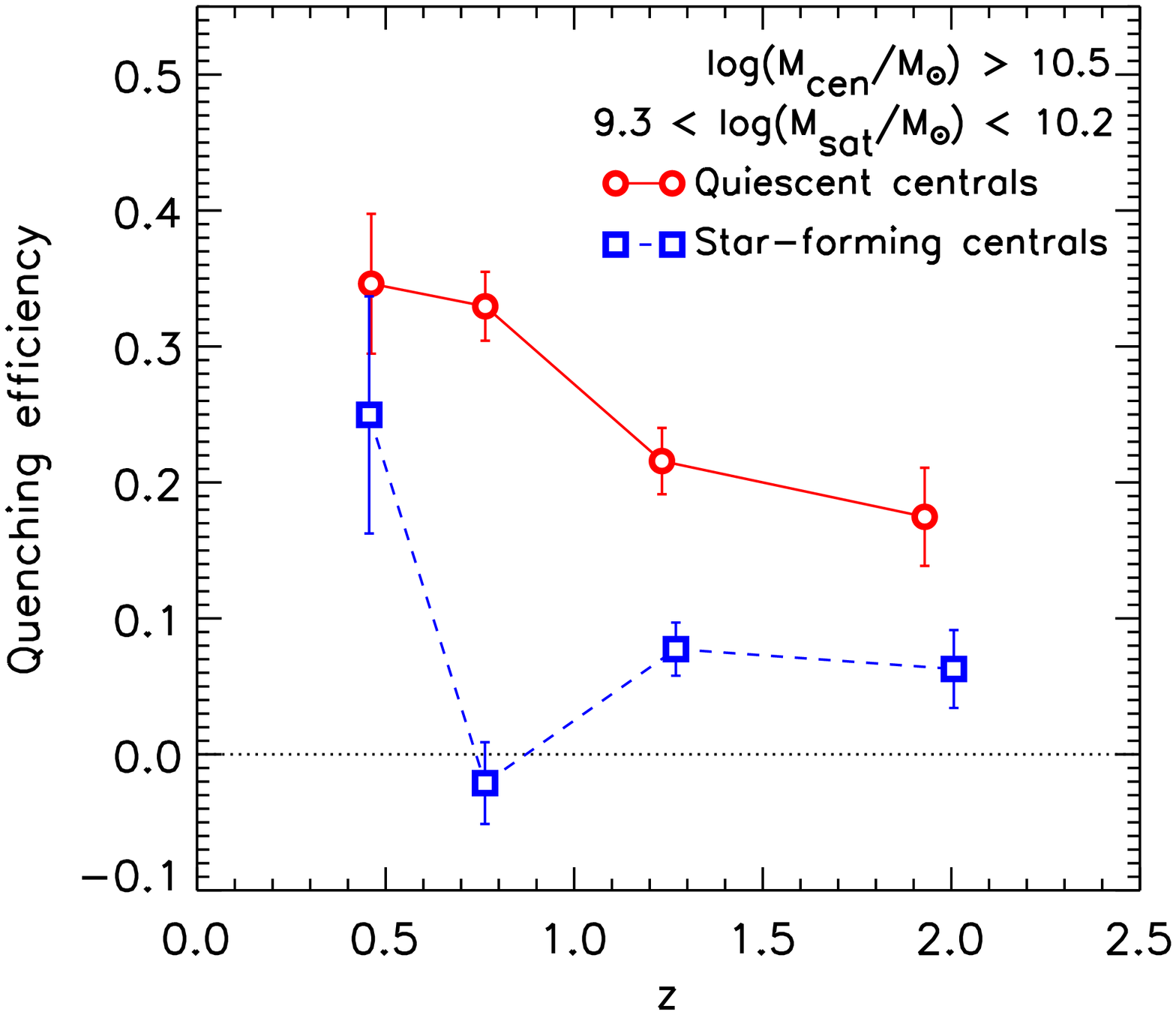}
\caption{The average quenching efficiency of satellites (\eq) 
with stellar mass of $9.3 < \log (M_\mathrm{sat}/\msol) < 10.2$ 
around central galaxies ($\log(M_{\unit{cen}}/\Msol)  > 10.5$) combining from three datasets.
The horizontal dotted line at  $\epsilon_{q,sat} = 0$ indicates no excess quenching of a satellite
 compared to mass-matched field samples.  Galactic conformity is evident as the higher 
quenching efficiency of satellites of quiescent centrals.  Satellites of star-forming centrals show 
low quenching efficiency for $z > 0.6$. For $z < 0.6$ there is evidence for elevated quenching of 
satellites of star-forming centrals (though still less than that for quiescent centrals).
Based on our bootstrap analysis, at $0.6 < z < 1.6$ the galactic conformity is significant at $3-4.5\sigma$, 
whereas the conformity at lowest and highest redshift is less significant.}
\label{fig:qeffcomballmsat}
\end{figure*}

\par We investigate how satellite quenching depends on the
star-formation activity of central galaxies by dividing our sample of
  central galaxies into subsamples that are star-forming and
  quiescent, where these labels correspond to galaxies with high and
  low sSFRs \citep{Williams2009,Papovich2012}, using their rest-frame $U-V$ and $V-J$ colors as
  illustrated in Figure~\ref{fig:uvjdiagram} and discussed in
  \S~\ref{sec:samplesel}. We then compute the quiescent fraction 
 (Equation~\ref{eq:fqsat}) and quenching efficiency of satellites (Equation~\ref{eq:eqsat}) for each subsample.
We use the evolving stellar-mass selection limit for satellites (\S ~\ref{sec:selsat}), and we apply the weighting factors to
match the stellar-mass distributions of star-forming and quiescent central galaxies (\S ~\ref{sec:matchmass}).
Error bars are estimated from a bootstrap resampling technique as described in \S ~\ref{sec:errorestimate}.
\par Figure~\ref{fig:qfrac4surveys} shows the satellite quiescent fraction for both quiescent and star-forming 
centrals from each dataset and each redshift bin.  At all redshifts, satellites of quiescent centrals have higher quiescent fractions 
compared to satellites of star-forming centrals.
Thus the phenomenon of galactic conformity can be seen in each of our datasets and in 
every redshift bin -- out to the highest redshifts probed by each dataset. This is one of the main 
conclusions of this paper.

Figure~\ref{fig:qeff4surveys} shows the satellite quenching
efficiency, which quantifies the excess quiescent fraction of
satellites compared to mass-matched field samples (see
Equation~\ref{eq:eqsat}).  This figure shows that satellites of both
quiescent and star-forming centrals have excess quenching
(i.e.,~positive quenching efficiency). The effect is most pronounced
for quiescent centrals, especially at $0.6 < z < 0.9$.
As discussed in\S~\ref{sec:intro}, there have been mixed results in the literature
regarding whether or not star-forming centrals can quench their
satellites; we find that they can. This suggests that the cause of
quenching in satellites is not tied directly to quenching in centrals,
i.e.,~that satellites can be quenched even when the central galaxy is
not. This is another primary conclusion of
this paper.


These conclusions can be seen more clearly in Figures~\ref{fig:qfraccomballmsat} and~\ref{fig:qeffcomballmsat},
where we show the quiescent fraction and quenching efficiency of satellites after combining the measurements
from the three datasets. Although satellites are quenched over time, we see evidence for galactic conformity
at all redshifts for centrals at fixed stellar mass. When the three fields are combined, there is significant, strong evidence that satellites around both star-forming and quiescent centrals have greater 
than zero quenching efficiencies: satellites have excess quenching above similar galaxies in the field regardless 
of the activity of their central galaxy.

\par At $0.6 < z < 0.9$ and $0.9 < z < 1.6$, there is high statistical
significance that satellites of quiescent centrals have a higher
quenching efficiency than satellites of star-forming centrals with
$p$ $\sim0.000001$ ($\simeq 4.5\sigma$) and
$p = 0.00021$ ($\simeq 3.5\sigma$), respectively. At $0.3 < z < 0.6$ the conformity signal is less significant ($p=0.088$, 1.4$\sigma$) and at $1.6 < z < 2.5$ there is no appreciable signal ($p=0.42$). 
%
%
Even though the survey volume is small and the statistical significance of the conformity signal is weak at $0.3 < z < 0.6$
(1.4$\sigma$ significance; see above and Figure~\ref{fig:qeffcomballmsat}), the signal is in line with what has been 
observed at even lower redshifts in SDSS (see \S~\ref{sec:intro}). Additionally, in the Appendix we show that the strength of the conformity signal at $0.3 < z < 0.6$ depends on the size of the aperture used to select satellites, where using different apertures can increase the conformity signal in this redshift bin, making it more in line with the SDSS results.
%

\par In the remainder of this paper we will continue exploring the dependence of satellite quenching by studying the quenching efficiency measured by combining all three datasets.


\subsection{Does Galactic Conformity Depend on the Central Mass?}
\label{sec:dependenceonmass}
\begin{figure*}
\epsscale{1.0}
\plotone{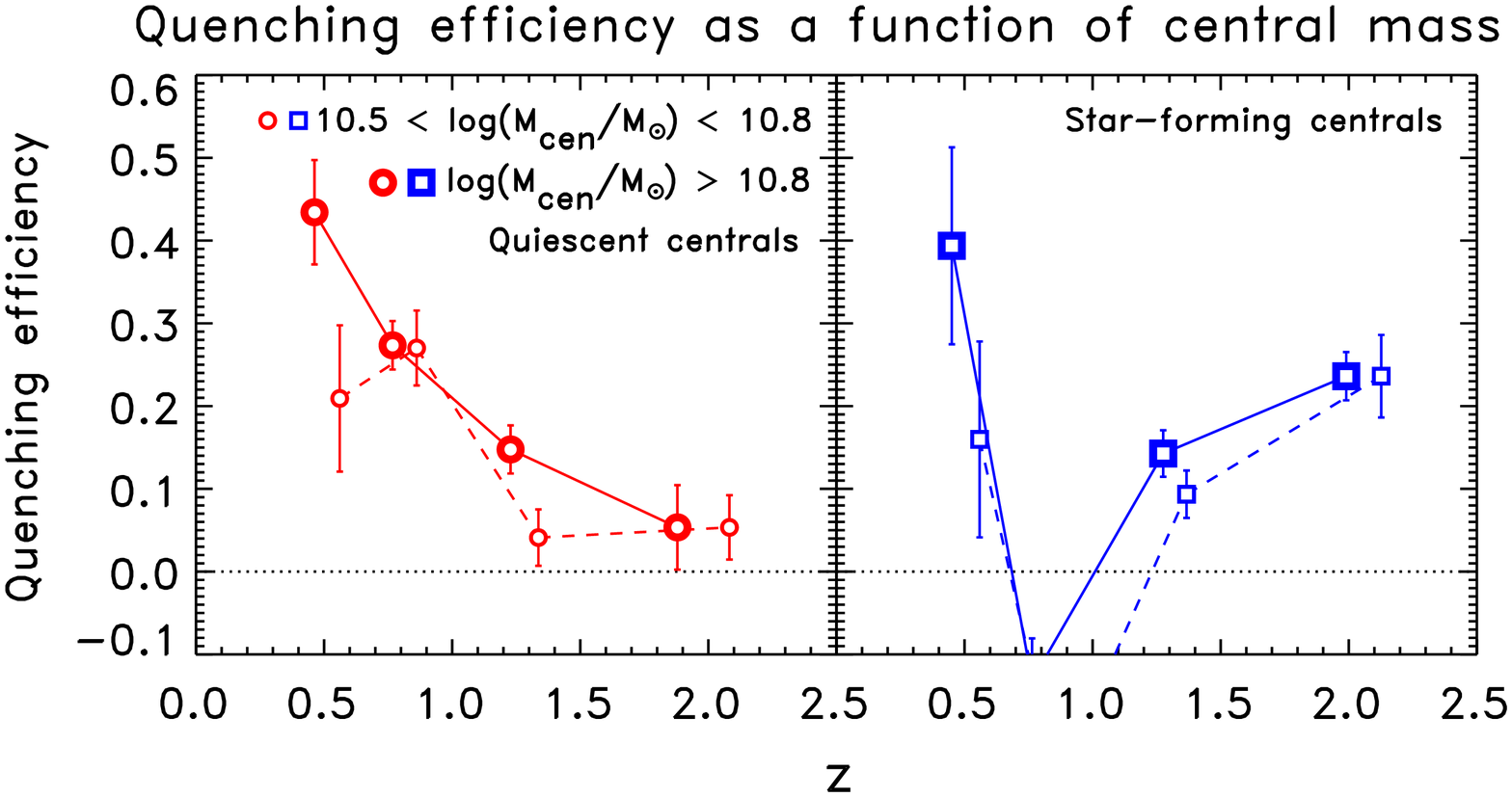}

\caption{\emph{Left:} Comparison between the average quenching efficiency of satellites 
($9.3 < \log (M_\mathrm{sat}/\msol) < 10.2$) around intermediate ($10.5 < \log(M_{\unit{cen}}/\Msol)  < 10.8$;
red solid line with small open circles, offset slightly for clarity) 
and high-mass ($\log(M_{\unit{cen}}/\Msol)  > 10.8$; red dash line with large open circles) quiescent centrals. 
\emph{Right:} Same as the left panel but for star-forming centrals.}
\label{fig:qcombmcen}
\end{figure*}

\par We divide our central galaxy sample
into two mass bins: $10.5 < \log(M_{\unit{cen}}/\Msol) < 10.8$ and 
$\log(M_{\unit{cen}}/\Msol) > 10.8$.
 We then recompute the quenching efficiency of satellites for each of  these subsamples 
to study the dependence of satellite quenching on the stellar mass of centrals
using the method described in \S~\ref{sec:selsat}.

\par As shown in Figure~\ref{fig:qcombmcen}, there is evidence for a dependence 
of satellite quenching on central mass for quiescent centrals at $0.3 < z < 0.6$ and 
$0.9 < z < 1.6$: satellites of more massive quiescent centrals at these redshifts
have a higher quenching efficiency. Similarly, for star-forming centrals, satellites of more 
massive centrals have a higher quenching efficiency at all redshifts, except at $1.6 < z < 2.5$. Figure~\ref{fig:qcombmcen} also shows that we observe the conformity 
of intermediate-mass centrals and high-mass centrals only at $0.6 < z < 0.9$.

\subsection{Does Galactic Conformity Depend on the Satellite Mass?}
\label{sec:dependenceonsatmass}

\par We divide our satellite galaxy sample with stellar masses in the range
$9.3 < \log(M_{\unit{sat}}/\Msol) < 10.2$  into two mass bins: 
$9.3 < \log(M_{\unit{sat}}/\Msol) < 9.8$ and 
$9.8 < \log(M_{\unit{sat}}/\Msol) > 10.2$. We then recompute the quenching 
efficiency of satellites for each of  these subsamples following the method
described in \S~\ref{sec:method}, and use an evolving stellar-mass limit for satellites (\S ~\ref{sec:selsat}).

\begin{figure*}
\epsscale{1.0}
\plotone{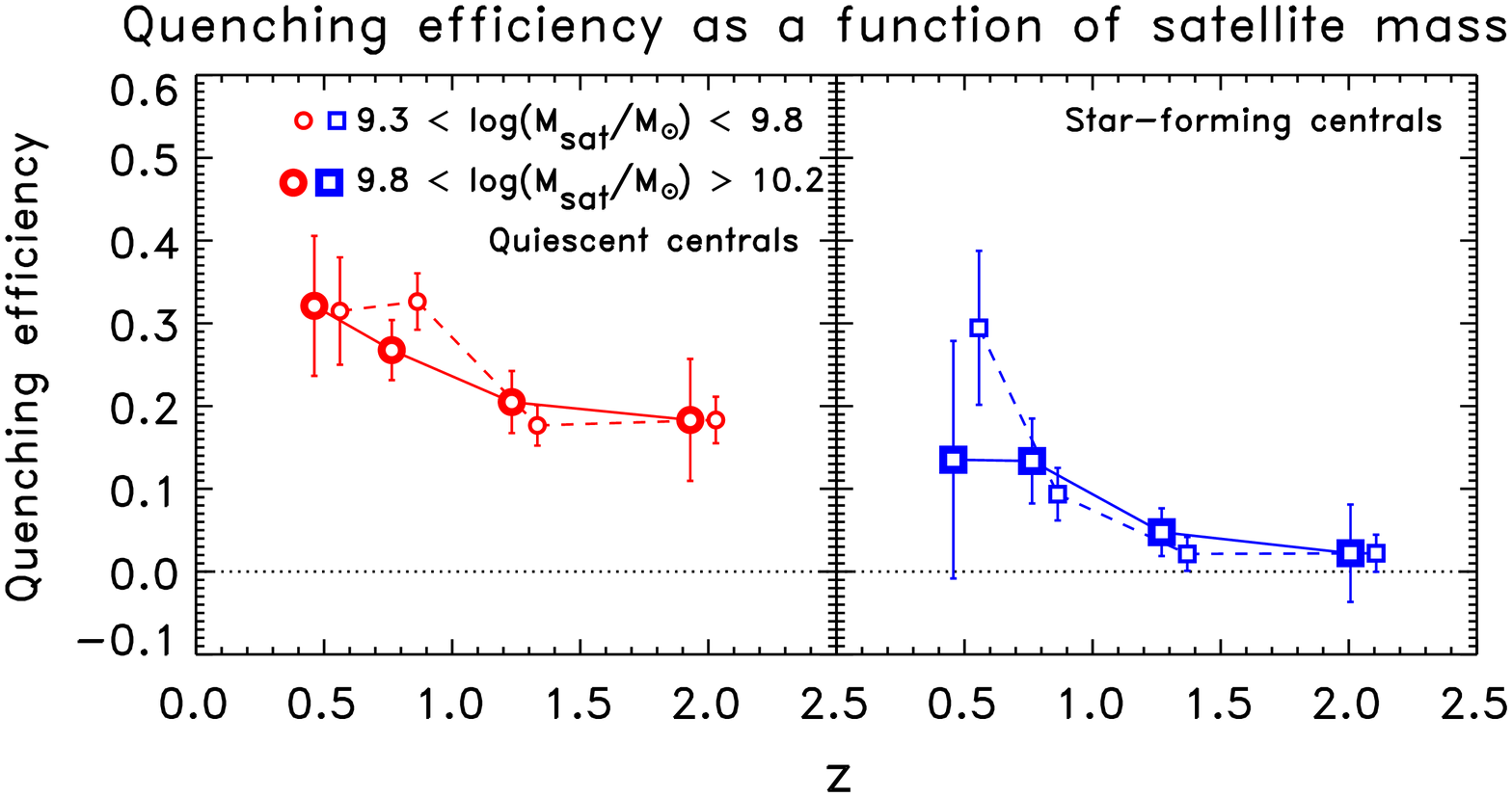}
\caption{\emph{Left:} Comparison between the average quenching efficiency of low-mass satellites ($9.3 < \log (M_\mathrm{sat}/\msol) < 9.8$; 
red dash line with small open circles) and high-mass satellites ($9.8 < \log (M_\mathrm{sat}/\msol) < 10.2$; red solid line with large open circles) around all quiescent centrals ($\log (M_\mathrm{cen}/\msol) > 10.5$). \emph{Right:} Same as the left panel but for satellites around
star-forming centrals.}
\label{fig:qcombmsat}
\end{figure*}

\par Figure~\ref{fig:qcombmsat} shows that there is no significant evidence
that the quenching efficiency depends on satellite mass for quiescent centrals.
%
%
The lack of a strong dependence of quenching on the mass of satellites is in agreement with the results from several studies \citep[e.g.,][]{VandenBosch2008,Peng2012,Quadri2012,Wetzel2013,Knobel2014}.
The galactic conformity signals for both low and high-mass satellites persist, except perhaps for 
%
low-mass satellites at $0.3 < z < 0.6$. 
We discuss the implications of these results in \S~\ref{sec:discussion}. 
\par A remaining question is how the conformity signal depends on the mass ratio of the central and satellite galaxies (rather than the absolute central and satellite stellar masses).  We attempted to investigate this effect by binning the sample by the stellar mass ratio between the satellites and centrals. However, this procedure severely limited the number of galaxies in the samples, such that we were unable to recover meaningful results. To study this effect we will require larger samples than are currently available.

\section{Discussion}
\label{sec:discussion}
In \S ~\ref{sec:satquenching} we showed that satellites around 
quiescent centrals have a higher quenching efficiency compared to satellites around 
star-forming centrals. This is galactic conformity, and it persists with high significance at 
intermediate redshift ($0.7 < z < 2.0$), and with a low level of significance at lower ($z\sim0.5$) and higher ($z\sim2.5$) redshift (Figure~\ref{fig:qfraccomballmsat} and~\ref{fig:qeffcomballmsat}).
In addition, the quenching efficiency of satellites aroung star-forming centrals is greater than zero
 indicating that satellites of star-forming centrals
are more quenched compared to background galaxies.
In this section we discuss the origin of the galactic conformity and the origin of the excess 
quenching of satellites of star-forming centrals.

\subsection{Does Halo Mass Drive Galactic Conformity?}
\label{sec:halomasstest}

Thus far we have investigated galactic conformity at fixed stellar mass, i.e.,~we have compared satellite quenching for samples of star-forming and quiescent centrals that have the same stellar mass distribution. However there is some observational evidence that, at fixed stellar mass, quiescent central galaxies occupy 
more massive halos than star-forming central galaxies 
\citep[e.g.,][]{Mandelbaum2006,More2011,Hartley2013,Phillips2014,Kawinwanichakij2014}.
This raises the possibility that, if satellite quenching is a function of halo mass, 
the observed conformity signal is due to a difference in halo mass rather than to 
a difference in star formation properties of central galaxies. As discussed in \S \ref{sec:intro}, there is some observational evidence that this is the case, although the results have been mixed. Ideally we would test this by matching the halo
masses of our quiescent and star-forming sample rather than matching the stellar
masses, but we lack halo mass estimates for the galaxies in our samples.


\par However, we can approximately match the halo masses of the star-forming and quiescent galaxy samples by matching their average number of satellites.  As we showed in a previous study \citep{Kawinwanichakij2014}, 
the number density of satellites around massive quiescent 
centrals ($\log(M_{\unit{cen}}/\Msol) > 10.8$) at $1 < z < 3$ from ZFOURGE/CANDELS is approximately twice as high as the number density of satellites  around star-forming centrals with the same stellar mass \citep[see also][]{Zheng2005}. We further argued in  \citeauthor{Kawinwanichakij2014}\ that the increase in satellites corresponds to a comparable increase in halo mass.
\begin{figure}
\epsscale{1.0}
\plotone{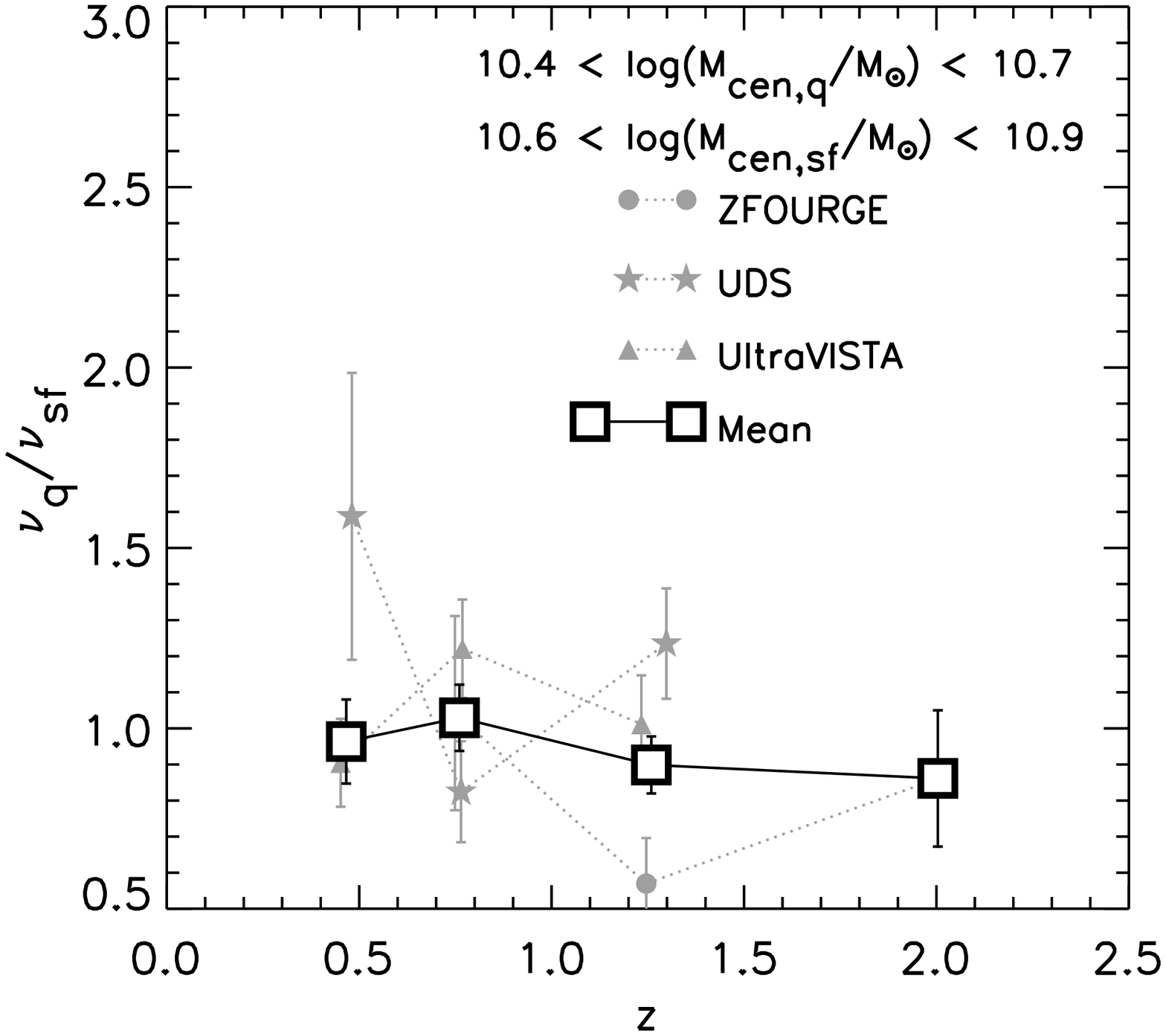}
\caption{Using the mean number of satellites per central galaxy to match approximately the halo masses of the quiescent and star-forming centrals.  The figure shows that the number of satellites per quiescent central ($\nu_{q}$) with $10.4 < \log(M_{\unit{cen}}/\Msol)  < 10.7$ is approximately the same as the
number of satellites per star-forming central ($\nu_{sf}$) with $10.6 < \log(M_{\unit{cen}}/\Msol)  < 10.9$. Assuming that the number of satellite scales with the halo mass of a central galaxy, this ratio implies that halo masses of our quiescent and star-forming here are roughly the same using these stellar mass ranges.}
\label{fig:qcombhaloA}
\end{figure}

\par We therefore make the assumption
that the mean number of satellites around our centrals is proportional to the halo mass. By selecting  samples of star-forming and quiescent centrals with the  same average number of satellites, we are able to select samples with approximately the same average halo mass and therefore  test if conformity can be explained by differences in halo mass. We define $\nu$ to be the average number of satellites per central for our sample. Figure~\ref{fig:qcombhaloA} shows that 
we can roughly match the number of satellites per central between star-forming ($\nu_{sf}$) 
and quiescent centrals ($\nu_{q}$) over our entire redshift range by selecting quiescent centrals with 
$10.4 < \log(M_{\unit{cen}}/\Msol) < 10.7$, and star-forming centrals with $10.6 < \log(M_{\unit{cen}}/\Msol) < 10.9$.
Therefore, we conclude that at fixed halo mass the quiescent centrals have stellar masses \textit{lower} by $\simeq$0.2 dex compared to the star-forming centrals.\footnote{If the halos that host quiescent centrals are older than the halos that host star-forming centrals, they may also have less substructure, and therefore will have fewer satellites at fixed halo mass. In this case our satellite-matching scheme would over-correct for differences in halo mass \citep[see also the discussion in][]{Hearin2015}} 

We have also tried estimating halo mass differences using total group stellar mass (i.e.,~the central mass plus the mass in detected satellites) rather than using the number of satellites \citep[e.g.][]{yang2007}. This leads to smaller halo mass differences between star-forming and quiescent centrals, and suggests that we are over-correcting for the halo mass by matching the number of satellites. Therefore, if anything our results should be conservative as we may be comparing satellites of star-forming centrals with slightly more massive halos to satellites of quiescent centrals with slightly less massive halos.

\begin{figure}
\epsscale{1.0}
\plotone{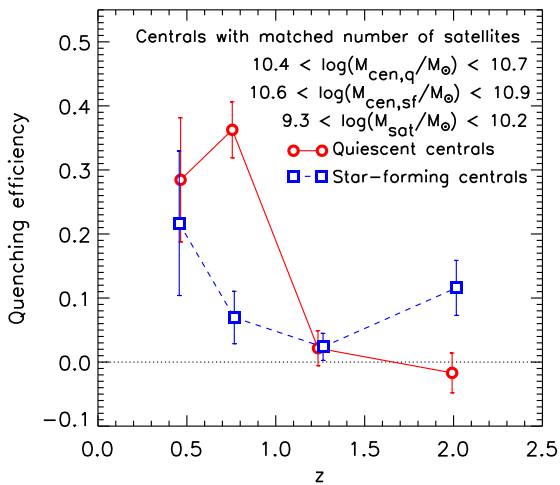}
\caption{The average quenching efficiency of satellites after approximately matching the halo masses of the quiescent and star-forming centrals using all three datasets.  The mean number of satellites is about equal for quiescent centrals with $10.4 < \log(M_{\unit{cen}}/\Msol)  < 10.7$ ,and star-forming centrals with $10.6 < \log(M_{\unit{cen}}/\Msol)  < 10.9$, implying they have approximately the same halo mass. Compared with Figure~\ref{fig:qeffcomballmsat}, the quenching efficiency of satellites of quiescent and star-forming centrals are about the same at all redshift (except at $0.6 < z < 0.9 $) after we matched the mean number of satellites  halo masses of quiescent and star-forming centrals. As discuss in \S~\ref{sec:discussion}, the galactic conformity observed in our galaxy sample is mainly driven by the halo mass. However, at $0.6 < z < 0.9$, the conformity is due to central galaxies being quiescent rather than just the halo mass of centrals.}
\label{fig:qcombhaloB}
\end{figure}

If satellite quenching was only a function of halo mass, with no residual correlation with the star-formation
activity of the central, then we would expect that the conformity signal would disappear when applying these different mass cuts.
Figure~\ref{fig:qcombhaloB} shows that, at $0.3 < z < 0.6$ and $z > 0.9$, the satellite quenching efficiency
around quiescent centrals and star-forming centrals are statistically equivalent when the
 mean halo mass of the star-forming centrals is about the same as the quiescent centrals.
The $p$-values derived from the bootstrap samples at $0.3 < z < 0.6$, $0.9 < z < 1.6$, and $1.6 < z < 2.5$, 
are $p_{\mathrm{}} = 0.34$, $p_{\mathrm{}} = 0.81$, and $p_{\mathrm{}} = 0.83$, implying that the satellites of quiescent 
and star-forming centrals are quenched equally at fixed halo mass. 
This suggests that, to within our uncertainties, halo-mass alone can account for galactic conformity at these redshifts.


\par However, halo mass appears not to account for all of the conformity signal at $0.6 < z < 0.9$.  
Figure~\ref{fig:qcombhaloB} shows that the conformity persists at $0.6 < z < 0.9$
even after we account for differences in the halo masses of the star-forming and quiescent centrals.
The significance of conformity at $0.6 < z < 0.9$, based on the bootstrap samples are
$p_{\mathrm{}} = 0.0004$ ($\simeq3.4\sigma$). 
Therefore, the observed galactic conformity at $0.6 < z < 0.9$ even at fixed halo mass implies that 
satellite quenching at this redshift range is related to the star-formation properties of the centrals in addition to just the halo mass.
Furthermore, the conformity signal at this redshift range at fixed halo mass is apparent in each of our datasets 
(ZFOURGE, UDS, and UltraVISTA), and is not driven by one individual field.

\subsection{Comparison to Previous Studies}

\par A number of studies analyzed the correlation between properties of 
satellites and their massive centrals (i.e., specific star-formation rate, colors, gas fraction) 
and the quiescent fraction of satellites in the local universe by utilizing the data from SDSS \citep{Kauffmann2010,Wang2012,Kauffmann2013,Knobel2014,Phillips2014,Phillips2015}. These authors have found that the quiescent fraction of satellites around quiescent centrals is higher 
than those of star-forming centrals. 

\par \citet{Phillips2014} reached the conclusion that massive satellites of
isolated star-forming centrals are indistinguishable from a field population, i.e.,~that satellite quenching does not occur in halos with star-forming centrals. This result, however, appears to be driven by 
their additional isolation criteria that allow no more than one satellite around their centrals. 
In a subsequent study, \citet{Phillips2015} demonstrated that star-forming centrals with two satellites have 
a non-zero satellite quenching efficiency. This is consistent with our result here: the higher quenching efficiency of satellites around star-forming centrals compared to the background galaxies for UDS and UltraVISTA at $0.3 < z < 0.6$, UltraVISTA at $0.6 < z < 0.9$, and all three surveys at higher redshift ranges. 

\par Our results extend trends from the lower-redshift to the higher-redshift Universe. 
As discussed in \S~1, earlier studies \citep[e.g.,][]{Weinmann2006,Knobel2014},  
have argued that the phenomenon of galactic conformity exists even after fixing the halo masses of the central galaxies -- although this conclusion is somewhat complicated by the results of \citet{Phillips2014,Phillips2015}. \cite{Wang2012} have demonstrated that conformity at fixed halo mass is present in the \citet{Guo2011} semi-analytic model, and suggest that this is because red centrals live in older halos, where satellite quenching is more efficient.
As shown in \S~\ref{sec:halomasstest}, our result (at least at $z<0.9$) is consistent with these studies in the sense that quiescent centrals have
a higher quiescent fraction compared to star-forming centrals, even after making a rough correction for the difference in halo mass.

\par The detection of galactic conformity out to $z\sim2$ was previously reported by \citet{Hartley2015} using 
an independent analysis of data from the UDS survey. Our analysis, which includes the UDS as well as the ZFOURGE and UltraVISTA 
surveys, bolsters this conclusion. We also find this conclusion persists to satellites of lower stellar mass 
($\log(M/\msol) = 9.3$). 

\par \citet{Hartley2015} also found that the quiescent fraction of satellites around 
star-forming galaxies is indistinguishable from the field population at all redshifts. 
When we restrict our analysis to the UDS sample only, we do find excess quenching
for the satellites of star-forming galaxies compared to the field in two of our three redshift bins, $0.3 < z < 0.9$ and 
$0.9 < z < 1.6$, in contrast to \citeauthor{Hartley2015}, but the significance is weak 
(Figures~\ref{fig:qfrac4surveys} and~\ref{fig:qeff4surveys}) and likely is a result of different analysis techniques and choice of aperture size.   
When we combine the UDS sample with our ZFOURGE and UltraVISTA samples, 
the signal becomes highly significant (Figure~\ref{fig:qeffcomballmsat}). \citeauthor{Hartley2015} also argued that halo mass alone is insufficient to account for all of the galaxy conformity signal by applying different stellar mass cuts to star-forming and quiescent centrals, as we have done here.



\par To summarize, when we take into account differences in sample selection and analysis,
the results from previous studies are consistent with ours in the sense that satellite galaxies are more quenched compared to the background galaxies, 
and the degree to which satellites have quenched is related to the star-formation activity 
of their central galaxies. The remaining differences in the quenching of satellites 
from our analysis and others may be a result of field-to-field (cosmic) variance, small number statistics, or
differences associated with the dataset, definition of quiescence, measurement techniques, and isolation criteria.  This emphasizes that systematics are still a significant contributor to the absolute measurements and studies of galactic conformity require multiple datasets and analysis techniques to understand the importance of these effects. 


\subsection{Physical Causes of Conformity}
\par In \S~\ref{sec:halomasstest} we suggested that the difference in
the halo mass of quiescent and star-forming centrals contributes to
the observed conformity signal, but there needs to be additional
mechanisms (at least at $0.6 < z < 0.9$). In this section, we discuss how halo mass can act as a driver of galactic conformity, and then we discuss additional possible origins of galactic conformity that may operate even at fixed mass.

\par It is generally argued that at a halo mass
$\sim 10^{12}$~$\Msol$, a halo of hot virialized gas is formed near
the virial radius \citep[e.g.,][]{White1978,Birnboim2003,Keres2005,Dekel2006}. This hot
halo shocks infalling cold gas to the virial temperature.  The hot
gas cools inefficiently, which may aid in reducing the star formation
in the central galaxy, but it is not expected to completely quench star
formation because of radiative cooling \citep{Birnboim2007}.  Additional
heating mechanisms have been proposed to prevent cooling of halo gas,
including AGN feedback \citep{Croton2006} or gravitational heating due
to clumpy accretion \citep{Birnboim2007,Dekel2007,Dekel2009}. The hot
gaseous halo surrounding quiescent centrals could also create an
environment which efficiently quenches satellite galaxies, either by
strangulation \citep{Larson1980} or ram pressure stripping
\citep{Gunn1972}, thereby causing galactic conformity. As more massive galaxies typically reside in more
massive host dark matter halos, the fact that we observe a positive
relationship between satellite quenching efficiency and the stellar
mass of star-forming centrals (Figure~\ref{fig:qcombmcen}) may reflect
the preference for more massive dark matter halos to harbor hot gas
coronas.


\par However there are several reasons to believe that there is more to the story. Observational \citep{Tumlinson2011,Churchill2013} and theoretical \citep{VandeVoort2011,Gabor2014} evidence suggests that the halos of quiescent central galaxies may have a significant cold gas component. In this study we have additionally found that satellites are quenched in the halos of star-forming centrals in excess of mass-matched field populations (Figure~\ref{fig:qeffcomballmsat}), suggesting that a hot gas halo does not always stifle star formation in the central itself. We also find that conformity persists even when we compare the quenched fractions of satellites of high-mass
star-forming centrals to lower-mass quiescent centrals (Figure~\ref{fig:qcombhaloB}), which provides some evidence that quenching is not simply a function of halo mass.

\par Even at fixed halo mass, there are several ways in which the environment within the halos of quiescent galaxies
may be more detrimental to star-formation in satellites. This could be due to a higher fraction of hot gas \citep[even at
  fixed halo mass;][]{Wang2012,Gabor2014}, which may be related to halo assembly history or to AGN feedback \citep{Croton2006}. It may also be due to tidal stripping or
harassment \citep{Farouki1981,Moore1996}. These effects could remove
gas from the satellite, where it could possibly contribute to the hot
halo and/or cool and accumulate on the central.


\par Even if satellites retain their own disk and (sub-) halo gas,
they will eventually exhaust that gas and may not accrete any more
gas. At low redshift, it has been argued that environmental processes
shut down star formation in satellites over a long timescale of $\sim
2-7$ Gyr in order to explain the distribution of satellite quiescent
fractions
\citep[e.g.,][]{Balogh2004,Finn2008,Weinmann2009,McGee2011,DeLucia2012,Haines2013,Wetzel2013}. The
reduced quenching of satellites at higher redshifts
(Figure~\ref{fig:qeffcomballmsat}) is therefore expected, since satellites will
not have had time to quench, however these timescales are
still too long to explain the existence of quenched satellites at these higher redshifts. This suggests that satellite quenching must proceed
more quickly at higher redshifts, as has been suggested previously \citep{Tinker2010,Quadri2012}. Faster gas depletion timescales at
higher redshift would also help to alleviate this problem.

\par In this regard, it is interesting to note that
\cite{Weinmann2010} showed that a model in which the diffuse gas is
stripped at the same rate that dark matter subhalos lose mass due to
tidal stripping can reproduce observations at low redshifts reasonably
well. This tidal stripping scenario has the attractive feature that it
operates more efficiently at higher redshifts, leading to shorter
quenching timescales. It also naturally explains the existence of
quenched satellites around star-forming centrals, since tidal
stripping takes place independent of the state of the halo gas and the
star-formation activity of central galaxies. Further work is required to determine whether tidal stripping can lead to conformity; it could be that the
conformity signal reflects earlier assembly time of certain halos, and
so the satellites have had more time to be stripped. 


\par Another fast-acting process that has not often been discussed in
this context is major merging. The violent dynamical environment of a
major merger may affect the sub-halos and satellite galaxies, however
future cosmological simulations into the behavior of the gas in
centrals and satellites in halos during major mergers would be needed
to test this scenario.


\par Satellite quenching and galactic conformity may also be related
to a class of effects due to the assembly history and large-scale
environment around dark matter halos, i.e.,~assembly bias. Because older halos will tend to have accreted their satellites long ago, those satellites will have had more time  to  lose  their  gas  supply  due  to  stripping  and exhaustion. Older halos are also expected to have higher concentration, which may also aid in tidal stripping of satellites. If older halos are \emph{also} more likely to host quiescent central galaxies \citep{Hearin2013,Hearin2014}, then this will naturally lead to conformity. Similar assembly bias effects are also relevant for observations of conformity beyond halo virial
radii (``2-halo conformity''). For instance, \citet{Kauffmann2013}
discuss the mass- and scale-dependence of conformity in the SDSS.  For low mass centrals, ($9.7 <
\log(M_{\unit{cen}}/\msol) < 10.5$), conformity extends out to
$\sim4~\unit{Mpc}$ around the centrals when they have low
star-formation rates or gas content.  This could be a result of a
correlation in the accretion rates of nearby halos \citep[as discussed
  by][]{Hearin2015}, but may also be due to large-scale heating of the intergalactic gas \citep[``preheating'';][]{Kauffmann2015}.\footnote{But see \citet{Paranjape2015},
  who suggest that the apparent large-scale 2-halo conformity may simply be
  due to the 1-halo conformity within the rare massive halos in
  the \citet{Kauffmann2013} sample.}  Similar large-scale correlations were also suggested by \cite{Quadri2012} as a possible way to explain the existence of a
star-formation density relation at $z\sim2$, and by \cite{Quadri2008}
and \cite{Tinker2010} as a way to help explain the strong clustering
of red galaxies at similar redshifts. 


\par \citet{Hearin2015} point out that large-scale 2-halo conformity will naturally lead to 1-halo conformity (which is what we are primarily measuring in this work) after the halos merge. These authors also find that 2-halo conformity due to assembly bias effects should vanish at $z > 1$. If 2-halo conformity were the \emph{only} cause of 1-halo conformity, then 1-halo conformity should decrease with redshift, and also vanish at $z > 1$. Our data show
significant 1-halo conformity to at least $z \sim 1.6$, in apparent contradiction with this
prediction.  It may be that 2-halo conformity extends to higher redshifts than predicted by \citet{Hearin2015}, or that 1-halo conformity is not simply caused by the correlated assembly histories of distinct dark matter halos at previous epochs. Additional large and deep datasets would be required to firmly establish or rule out the existence of 1-halo and 2-halo conformity at these and higher redshifts.

\par Finally we note that, if halo age or recent assembly history are important causes of (either 1-halo or 2-halo) conformity, then this requires that the baryonic physics of star formation and quenching are sensitive to halo assembly history. As mentioned above, this seems obvious in the case of satellite quenching: satellites with early accretion times are more likely to be quenched. However it is less obvious that quenching of central galaxies should be strongly tied to halo accretion rate. If infalling gas is shock-heated and is added to a hot gaseous halo \citep[as is generally expected at $z<2$; e.g.][]{Dekel2006}, rather than penetrating to the central regions, then it is not clear that the central star formation should couple strongly to the halo accretion rate. Conversely, even halos with low accretion rates are expected to contain significant hot gas components, which can in principle provide fuel for star formation. Hydrodynamic simulations are necessary to investigate whether low halo accretion rates can be a significant factor in the quenching of central galaxies over the redshift range where conformity is now known to exist; \citet{Feldmann2015} have recently demonstrated this at $z > 2$, but their simulations do not extend to lower redshift.

\section{Summary}

We have studied the quiescent fraction (\fq) and quenching efficiency (\eq) of satellites around
star-forming and quiescent central galaxies with $\log(M_\mathrm{cen}/\mathrm{M}_\odot) > 10.5$
 at $0.3 < z < 2.5$. We use data from three different deep near-IR surveys ZFOURGE/CANDELS,
UDS, and UltraVISTA that span different ranges of depth and area in order to achieve
significant volume at lower redshifts as well as sufficient depth for high redshift measurements. 
The deep near-IR data allow us to select satellites down to $\log(M/\msol)>9.3$ at $z<2.5$. 
The main conclusions of this work are the following:

\begin{itemize}

\item We find that satellite galaxies, $9.3 < \log(M_{\unit{sat}}/\msol) < 10.2 $
at $0.3 < z < 2.5$ are more quenched compared to mass-matched samples of field galaxies.

\item Galactic conformity exists at $0.3 < z < 2.5$: while the satellites of star-forming central galaxies are quenched in excess of field galaxies, the satellites of quiescent centrals are quenched at an even higher rate. There is a strong conformity signal at $0.6 < z < 0.9$ ($4.5\sigma$) and at $0.9 < z < 1.6$ ($3.5\sigma$), whereas the conformity in our lowest and highest redshifts bins, $0.3 < z < 0.6$ and $1.6 < z < 2.5$, is less significant. This may be a real physical effect, or may be due to insufficient statistics. Regardless, conformity is not a recent effect, but has been present for a significant fraction of the age of the universe --- conformity may even be as old as satellite-quenching itself.

\item A comparison between the quenching efficiency of intermediate-mass centrals ($10.5 < \log(M_{\unit{cen}}/\msol) < 10.8$) and high-mass centrals 
($\log(M_{\unit{cen}}/\msol) > 10.8$) indicates that satellite quenching depends on the stellar mass of the central, in that satellites around more massive centrals have a higher quenching efficiency. This appears to be true for both star-forming and quiescent centrals.

\item The existence of galactic conformity is observed for both low mass ($9.3 < \log(M_{\unit{sat}}/\msol) < 9.8$) 
and high mass satellites ($9.8 < \log(M_{\unit{sat}}/\msol) < 10.2$) 
around centrals of all masses and redshifts (with the possible exception of the
highest mass satellites at the highest redshifts $1.6 < z < 2.5$, 
and the lowest mass satellites at the lowest redshifts $0.3 < z < 0.6$, where our statistics are poorer).
There is no significant evidence that satellite quenching depends on the stellar mass of the satellites.

\item We test if galactic conformity is due to a difference in the
  typical halo mass of star-forming and quiescent centrals by
  selecting star-forming centrals with $\sim0.2$ dex higher stellar
  mass. This difference should be enough to eliminate any difference
  in halo mass between our quiescent and star-forming samples. From
  this test we find that the difference in halo mass can explain most
  of the conformity signal in our data.  However, there still remains
  evidence for conformity, particularly at $0.6 < z < 0.9$. This suggests that
  satellite quenching is connected to the star-formation properties of
  the central, beyond the mass of the halo.

\item While halo mass may be a significant (even dominant) driver of conformity, it does not appear to explain all of the conformity signal.  We have discussed other physical effects that may account for the existence and evolution of the conformity signal, including hot gas halos, feedback effects, halo assembly history, and large-scale environment -- and we have discussed some of the issues involved with these explanations.


\end{itemize}

\acknowledgements

We wish to thank our collaborators in the \zfourge\ and CANDELS
teams for their dedication and assistance, without which this work
would not have been possible. We also wish to thank Andrew Wetzel, Duncan Campbell for valuable comments and feedback. 
We also thank the anonymous referee for a very constructive and helpful report. Australian access to the 
Magellan Telescopes was supported through the National Collaborative 
Research Infrastructure Strategy of the Australian Federal Government.
This work is supported by the National Science Foundation
through grant AST 1009707 and AST 1413317. This work is based on observations taken
by the CANDELS Multi-Cycle Treasury Program with the NASA/ESA HST, which is operated by the
Association of Universities for Research in Astronomy, Inc., under
NASA contract NAS5-26555.  This work is supported by HST program
number GO-12060.  Support for Program number GO-12060 was provided by
NASA through a grant from the Space Telescope Science Institute, which
is operated by the Association of Universities for Research in
Astronomy, Incorporated, under NASA contract NAS5-26555.  
We acknowledge generous support from the
Texas A\&M University and the George P.\ and Cynthia Woods Institute
for Fundamental Physics and Astronomy.
GGK acknowledges the support of the Australian Research Council through the award of a Future Fellowship (FT140100933).
We acknowledge support from NL-NWO Spinoza.
\begin{appendix}
\section{The Dependence of Satellite Quenching on the Aperture Size}
\label{sec:appendix}
\begin{figure}
\epsscale{1.0}
\plotone{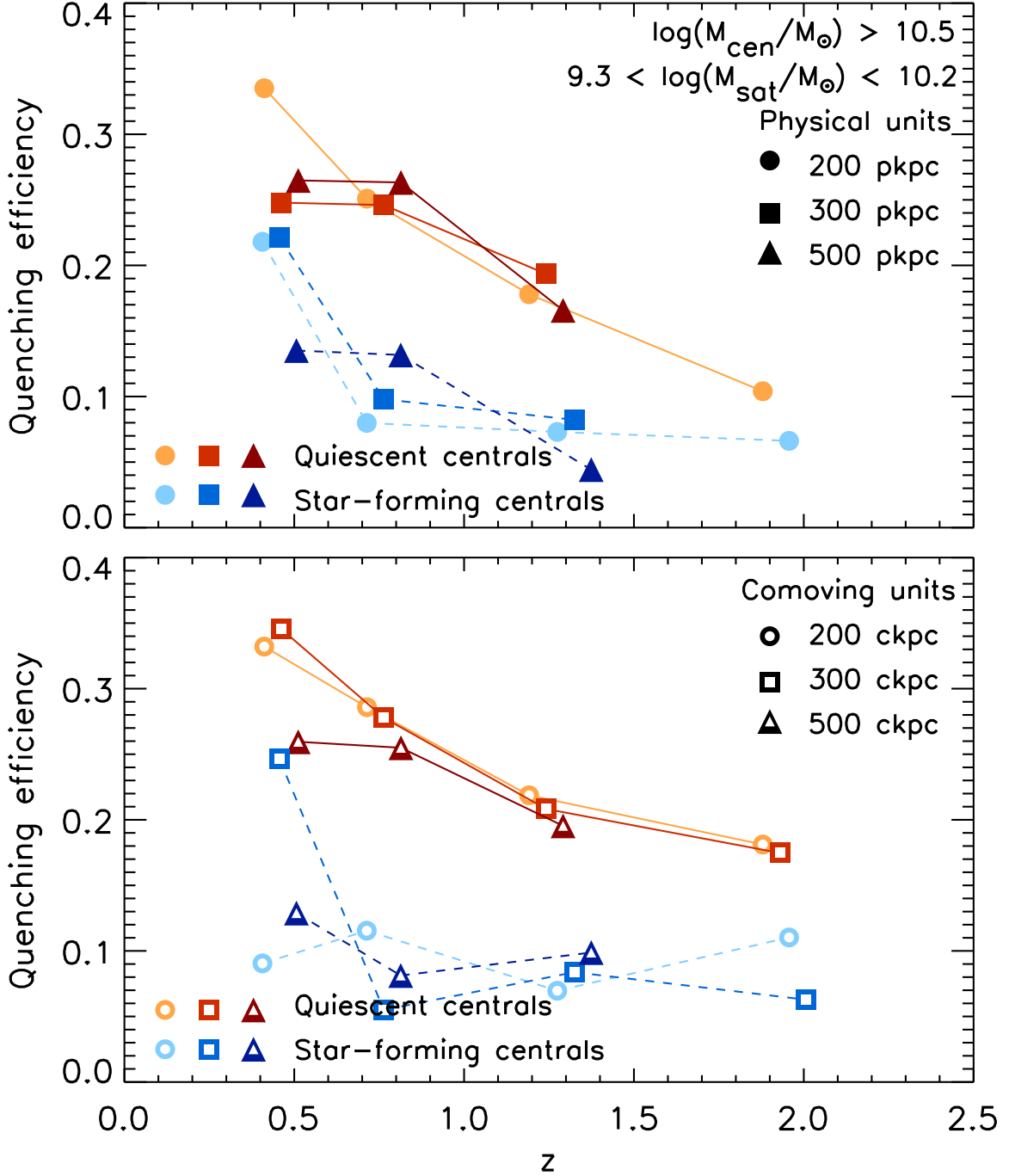}

\caption{
The satellite quenching efficiency as a function of redshift and central galaxy type, where the satellites are identified within different aperture sizes around the central galaxies. The satellites have stellar mass $9.3 < \log (M_\mathrm{sat}/\msol) < 10.2$ 
and the central galaxies have $\log(M_{\unit{cen}}/\Msol)  > 10.5$. The top and bottom panels show the results in apertures that have fixed radii in physical and comoving units, respectively. (In the highest redshift bin, $1.6 < z < 2.5$ the 300~pkpc, 500~pkpc,and 500~ckpc probe a significant portion of the image: at $z\sim 2$, 500 pkpc corresponds to $0\farcm5$, making the measurements intractable, and we do not include them here.)  There is no strong dependence on the strength of the quenching efficiency with the choice of aperture radius, with the possible exception of the $0.3 < z < 0.6$ bin. Uncertainties have been suppressed for clarity.}
\label{fig:apertures}
\end{figure}

The comparison of satellite galaxy quenching and galactic conformity in the literature is complicated because different studies use a wide range of aperture sizes within which to identify satellites \citep[e.g.,][]{Wang2012,Tal2014a,Phillips2014,Phillips2015,Hartley2015}. The primary results in this study are based on a 300~ckpc aperture, but in this appendix we show how the use of different aperture sizes affects the quenching efficiencies.

We recomputed the quenching efficiencies of satellites for the central and star-forming galaxy samples in each redshift bin using different aperture sizes, including both comoving and physical aperture radii: 200~ckpc, 300~ckpc, 500~ckpc, 200~pkpc, 300~pkpc, and 500~pkpc.
The results are shown in Figure~\ref{fig:apertures} and are tabulated in Table~\ref{table:qfracapersizes}. 

The observed conformity signal does not depend strongly on the choice of aperture.   At redshifts $0.6 < z < 2.5$ there is no significant dependence on the quenching efficiency on aperture.  The biggest difference is apparent in our lowest redshift bin, $0.3 < z < 0.6$, where we see that the quenching efficiency of satellites of star-forming galaxies can be reduced, thereby increasing the strength of galactic conformity.  However, these are still within the errors (see Figure~7 and Table~\ref{table:qfracapersizes}).

%
%

\begin{deluxetable*}{ccccccc}
\tabletypesize{\footnotesize}
\tablecolumns{5} 
\tablewidth{0pt}
 \tablecaption{Quiescent fractions ($f_{\unit{q}}$) and quenching efficiency ($\epsilon_{\unit{q}}$) of satellites of quiescent and star-forming centrals measured in different apertures size
 \label{table:qfracapersizes}}
 \tablehead{
 \colhead{Stellar mass range} \vspace{-0.1cm}& \colhead{Redshift} &  \colhead{Aperture size} & \colhead{$f_{\unit{q},\mathrm{Quiescent}}$} &
 \colhead{$f_{\unit{q},\mathrm{Star-forming}}$} &
 \colhead{$\epsilon_{\unit{q},\mathrm{Quiescent}}$} &
 \colhead{$\epsilon_{\unit{q},\mathrm{Star-forming}}$}\\ 
\vspace{0.1cm}} 
 \startdata 
 \cutinhead{\bf{Central mass:} $\log(M_{\mathrm{cen}}/\Msol) > 10.5$}

           \bf{Satellite mass:}  $\log(M_{\mathrm{sat}}/\Msol) = 9.3-10.2$   &  $ 0.3 < z < 0.6 $ &200~ckpc &0.43&0.23 & 0.33 & 0.09 \\ 
                     &                             & 300~ckpc & $0.45\pm0.04$ & $0.35\pm0.07$ &$0.35\pm0.05$& $0.25\pm0.09$ \\ 
                     &                             & 500~ckpc & 0.37 & 0.26 & 0.26 & 0.13 \\
                     &                             & 200~pkpc & 0.43  &0.33 & 0.34&0.22 \\ 
                      &                             &300~pkpc & 0.36  & 0.33 & 0.25& 0.22\\ 
                      &                             &500~pkpc & 0.37  & 0.26 & 0.26 &0.14 \\ \\
           \cline{2-7}     
                      &                           &                    &         &   & &   \\
                     &  $ 0.6 < z < 0.9 $         & 200~ckpc           & 0.40 & 0.25 &0.29 &0.12\\ 
                     &				               & 300~ckpc & $0.44\pm0.02$ & $0.16\pm0.03$& $0.33\pm0.03$ & $-0.02\pm0.03$\\
                     &                             & 500~ckpc   &0.37  & 0.22 & 0.25&0.08\\ 
                     &                             & 200~pkpc   & 0.36  & 0.21 & 0.25 & 0.08\\ 
                     &                             & 300~pkpc   & 0.36  &0.22  & 0.25 &0.10\\ 
                     &                             & 500~pkpc   & 0.37 & 0.26  &0.26& 0.13\\  \\
             \cline{2-7}  
              &                           &                    &         &  & &    \\
                      & $ 0.9 < z < 1.6 $         &200~ckpc	& 0.31& 0.18 & 0.22 & 0.07  \\
                      &                           & 300~ckpc  & $0.30\pm0.02$&$0.18\pm0.02$&$0.22\pm0.02$ & $0.08\pm0.02$ \\
                      &                           & 500~ckpc   & 0.29   & 0.20 & 0.20 &0.10 \\ 
                      &                           & 200~pkpc   & 0.27   & 0.18 & 0.18 & 0.07 \\ 
                      &                           & 300~pkpc   & 0.29   & 0.19 &0.19 & 0.08 \\ 
                      &                           & 500~pkpc   & 0.26   & 0.16 &0.17 & 0.04 \\ \\
               \cline{2-7}  
                      &                           &                &         &    & &  \\
                      & $ 1.6 < z < 2.5 $         & 200~ckpc      &0.23 &0.17& 0.18 & 0.11 \\
                      &                           & 300~ckpc   & $0.22\pm0.03$  & $0.12\pm0.03$&$0.17\pm0.04$ & $0.06\pm0.03$ \\
                      &                           & 200~pkpc   &0.16  & 0.12 & 0.10 & 0.07\\ 
                      &                           & 300~pkpc   &0.10   & 0.11 &0.04 & 0.05\\ \\
                    &                           	&                   &         &  & & \\ 

 \enddata
\vspace{-0.3cm}
 \tablecomments{The uncertainties of quiescent fractions and quenching efficiencies of satellites measured in aperture radii of 200~ckpc, 500~ckpc, 200~pkpc, 300~pkpc, and 500~pkpc are not shown and are assumed to be the same as the uncertainties measured in 300~ckpc aperture radius. }

\end{deluxetable*}

\end{appendix}

\bibliography{references}

\end{document}